\documentclass[11pt, letterpaper,  onecolumn]{IEEEtran}
\ifCLASSINFOpdf
\else
\fi
\usepackage{graphicx}
\usepackage{amsmath}
\usepackage{amssymb}
\usepackage{graphics}
\usepackage{latexsym}
\usepackage[dvips]{epsfig}
\usepackage[nospace]{cite}
\usepackage{subfigure}
\usepackage{setspace}
\usepackage{tabularx}

\newtheorem{Definition}{Definition}
\newtheorem{Lemma}{Lemma}

\newtheorem{Proposition}[Lemma]{Proposition}
\newtheorem{Theorem}{Theorem}

\usepackage[ colorlinks = true,
             linkcolor = blue,
             urlcolor  = blue,
             citecolor = red,
             anchorcolor = green,
]{hyperref}

\allowdisplaybreaks[1]

\def\Pr{{\mathrm{Pr}}}
\def\E{{\mathrm E}}
\def\Var{{\mathrm {Var}}}
\def\Cov{{\mathrm {Cov}}}

\begin{document}
\title{Asymptotic Expansions for Gaussian Channels with Feedback under a Peak Power Constraint}

%\author{Silas~L.~Fong and Raymond~W.~Yeung,~\IEEEmembership{Fellow,~IEEE}% <-this % stops a space
%\thanks{Silas~L.~Fong is with the Department of Electronic Engineering, City University of Hong Kong, Kowloon, Hong Kong (e-mail: lhfong5@ie.cuhk.edu.hk).}% <-this % stops a space
%\thanks{Raymond~W.~Yeung is with the Institute of Network Coding and the Department
%of Information Engineering, The Chinese University of Hong Kong,
%N.T., Hong Kong (e-mail: whyeung@ie.cuhk.edu.hk).}}% <-this % stops a space
%%\thanks{Manuscript received April 19, 2005; revised January 11, 2007.}}

\author{Silas~L.~Fong and Vincent~Y.~F.~Tan% <-this % stops a space
\thanks{Silas~L.~Fong and Vincent~Y.~F.~Tan are with the Department of Electrical and Computer Engineering, National University of Singapore (NUS), Singapore (e-mail: \texttt{\{silas\_fong,vtan\}@nus.edu.sg}). Vincent~Y.~F.~Tan  is also with the Department of Mathematics, NUS. }}% <-this % stops a space
\maketitle
\flushbottom

\begin{abstract}
This paper investigates the asymptotic expansion for the size of block codes defined for the additive white Gaussian noise (AWGN) channel with feedback under the following setting: A peak power constraint is imposed on every transmitted codeword, and the average error probability of decoding the transmitted message is non-vanishing as the blocklength increases. It is well-known that the presence of feedback does not increase the first-order asymptotics (i.e., capacity) in the asymptotic expansion for the AWGN channel. The main contribution of this paper is a self-contained proof of an upper bound on the asymptotic expansion for the AWGN channel with feedback. Combined  with existing achievability results for the AWGN channel, our result implies that the presence of feedback does not improve the second- and third-order asymptotics. An auxiliary contribution is a proof of the strong converse for the parallel Gaussian channels with feedback under a peak power constraint.
%\boldmath
\end{abstract}

\begin{IEEEkeywords}
%apacity region, cut-set outer bound, discrete memoryless network (AWGN channel), strong converse.
AWGN channel, Feedback, Asymptotic expansion, Second-order asymptotics, Parallel Gaussian channels
\end{IEEEkeywords}

% \fi
\IEEEpeerreviewmaketitle

\section{Introduction} \label{Introduction}
The additive white Gaussian noise (AWGN) channel is one in which at each discrete time $k \in \{1,2,\ldots, n\}$, the output $Y_k$ is the sum of the input $X_k$ and a Gaussian random variable $Z_k$ that represents additive noise. The collection of the noise random variables $\{Z_k\}_{k\in \{1,\ldots, n\}}$ is assumed to be independent and identically distributed (i.i.d.). The inputs are also power limited, which means that $\sum_{k=1}^n X_k^2 \le nP$ with probability 1 where $P>0$ is the permissible power, i.e., a peak power constraint. If we would like to transmit a uniformly distributed message $W\in \{1,2,\ldots, \lceil 2^{nR} \rceil\}$ across this channel, it was shown by Shannon \cite{Shannon48} the maximum rate of communication $R$ or the {\em capacity} is
\begin{equation}
\mathrm{C}(P) \triangleq \frac{1}{2}\log(1+P) \qquad\mbox{bits per channel use}. \label{defCP}
\end{equation}
In other words, if $M^*(n,\varepsilon,P)$ designates the maximum number of messages that can be transmitted over $n$ uses of an AWGN channel with permissible power $P$ and average error probability $\varepsilon$, one has
\begin{equation*}
\lim_{\varepsilon\downarrow 0}\liminf_{n\to\infty}\frac{1}{n}\log M^*(n,\varepsilon,P)=\mathrm{C}(P).
\end{equation*}
In fact, the strong converse was shown by Shannon in~\cite{Sha59b} (also see Yoshihara~\cite{Yoshihara} and Wolfowitz~\cite{Wolfowitz}) and   so we have
\begin{equation*}
\lim_{n\to\infty}\frac{1}{n}\log M^*(n,\varepsilon,P)=\mathrm{C}(P)
\end{equation*}
for every $\varepsilon\in (0,1)$.

{\em Feedback}, which is the focus of the current paper, is known to simplify coding schemes and   improves the performance of communication systems in many scenarios. See \cite[Chapter 17]{elgamal} for a thorough discussion of the benefits of feedback in single- and multi-user information theory. When feedback is allowed, each input symbol $X_k$ depends not only on the transmitted message $W$ but also the vector of channel outputs up to and including time $k-1$, i.e., the symbols  $Y^{k-1}= (Y_1, \ldots, Y_{k-1})$. For {\em memoryless} AWGN channels, it is known that feedback does not increase the capacity of the channel, i.e., the feedback capacity remains at $\mathrm{C}(P)$.  This follows from  a seminal result by   Shannon~\cite{Sha56} in which he proved that noiseless feedback does not increase the capacity of memoryless channels.

In this paper, we are interested in analyzing the performance of the AWGN channel with feedback under the constraint that the average error probability in decoding the transmitted message is non-vanishing, i.e., bounded above by a constant $\varepsilon\in (0,1)$. In the absence of feedback, it is known from Polyanskiy-Poor-Verd\'u \cite[Theorem 54, Eq.\ (294)]{PPV10} and Tan-Tomamichel \cite[Theorem 1]{TanTom13a} that
\begin{equation}
\log M^*(n,\varepsilon,P)=n\mathrm{C}(P) + \sqrt{n\mathrm{V}(P)} \Phi^{-1} (\varepsilon) + \frac{1}{2}\log n + O(1) \label{eqn:asymp_expans}
\end{equation}
where
\begin{equation}
\mathrm{V}(P) \triangleq  \frac{P(P+2) (\log e)^2}{2(P+1)^2}\qquad\mbox{bits}^2 \mbox{ per channel use} \label{defVP}
\end{equation}
is known as the Gaussian {\em dispersion} function and $\Phi^{-1}$ is the inverse of the cumulative distribution function for the standard Gaussian distribution. See Hayashi's work~\cite{Hayashi09} for a proof of \eqref{eqn:asymp_expans} without the  third-order $\frac{1}{2}\log n +O(1)$ term.

\subsection{Main Contributions}
A natural question then arises. In the presence of feedback, what is the analogue of the asymptotic expansion in \eqref{eqn:asymp_expans}? Let $M_{\mathrm{fb}}^*(n,\varepsilon,P)$ be the maximum number of codewords that can be transmitted through $n$ uses of the channel when each input symbol $X_k$ is allowed to depend on $(W,Y^{k-1})$. Clearly, $M_{\mathrm{fb}}^*(n,\varepsilon,P)\ge M^*(n,\varepsilon,P)$ for all choices of the parameters $(n,\varepsilon,P)$ (because the code can simply ignore the fed back symbols $Y^{k-1}$).  In this work, our main contribution is a conceptually simple, concise and self-contained proof that the asymptotic expansion in \eqref{eqn:asymp_expans} remains unchanged, i.e.,
\begin{equation}
\log M_{\mathrm{fb}}^*(n,\varepsilon,P)=n\mathrm{C}(P) + \sqrt{n\mathrm{V}(P)} \Phi^{-1} (\varepsilon) + \frac{1}{2}\log n + O(1) .\label{eqn:asymp_expans_fb}
\end{equation}
This means that, up to the third-order term in the asymptotic expansion of $\log M_{\mathrm{fb}}^*(n,\varepsilon,P)$, full feedback from the output of the channel to the encoder does not increase the number of codewords transmissible over the channel.

%This is somewhat surprising  (at least to the authors) given that the error probability performance improves greatly in the presence of feedback for fixed rates below capacity~\cite{Schalkwijk,Schalkwijk1,  Shayevitz}.

As an auxiliary contribution, we investigate the parallel Gaussian channels with feedback under a peak power constraint and prove an upper bound for the second-order asymptotics. This establishes the strong converse for this channel, which (to the best of the authors' knowledge) was not known previously.
%\subsection{Main Contribution}

\subsection{Related Work}
Our work is inspired by Altu\u{g} and Wagner's recent study of the fundamental limits of communication over discrete memoryless channels (DMCs) with feedback \cite{AW14}. In their work,  Altu\u{g} and Wagner showed \cite[Theorem 1]{AW14} that for some classes of DMCs whose capacity-achieving input distributions are not unique (and in particular the minimum and maximum conditional information variances differ), the second-order asymptotics improves in the presence of feedback compared to the no-feedback scenario. They also showed \cite[Theorem 2]{AW14} that feedback does not improve the second-order asymptotics for DMCs $p_{Y|X}$ if the conditional variance of the log-likelihood ratio $\log \frac{p_{Y|X}(Y|x) }{q^*(Y)}$, where $q^*$ is the unique capacity-achieving output distribution, does not depend on the input $x$. Such DMCs include the class of weakly-input symmetric DMCs initially studied by Polyanskiy-Poor-Verd\'u~\cite{PPV11b}. Our contribution is similar in spirit to \cite[Theorem 2]{AW14}. However, we note that the proof technique used by Altu\u{g} and Wagner requires the use of a sophisticated Berry-Ess\'een-type result for bounded martingale difference sequences~\cite{Machkouri}. Our technique for the AWGN channel is conceptually simpler. We prove that a sum of random variables that naturally appears in the non-asymptotic analysis of the AWGN channel with feedback has the same distribution as the sum of i.i.d.\ random variables, thus facilitating the use of the usual Berry-Ess\'een theorem~\cite[Theorem 2 in Section XVI.5]{feller}. We prove this equivalence between the distributions by using moment generating functions.

%Our result in  \eqref{eqn:asymp_expans_fb} is somewhat expected in light of the fact that
In another line of work, for rates below capacity $\mathrm{C}(P)$, Pinsker~\cite{Pin68} showed that for fixed-length block codes on Gaussian channels, the use of feedback cannot improve the exponent over the sphere-packing bound under the peak power constraint. This is in contrast to the case where an expected average power constraint is imposed under which Schalkwijk and Kailath \cite{Schalkwijk,Schalkwijk1} showed in a series of celebrated works that the error probability decays doubly exponentially fast for rates below $\mathrm{C}(P)$. The work contained herein {\em only} considers the peak power constraint. We leave the analysis of the asymptotic expansion for $\log M_{\mathrm{fb}}^*(n,\varepsilon,P)$ under the expected average power constraint for future work.

%In another line of work, Schalkwijk and Kailath \cite{Schalkwijk} and Schalkwijk \cite{Schalkwijk1} showed for the AWGN channel with feedback that for fixed rates below capacity, the error probability decays doubly exponentially fast. This coding strategy now known as {\em posterior matching} has been and studied more extensively for a much wider class of channels by Shayevitz and Feder~\cite{Shayevitz}. It has also been showed by Pinsker \cite{Pin68}, Zigangirov~\cite{Zigangirov70} and Kramer~\cite{Kramer69} that the probability of error can be made to decay as fast as an arbitrary level of nested exponentials.  The fact that the asymptotic expansion of $\log M^*(n,\varepsilon,P)$  (up to the third-order term) is unaffected by the presence of feedback lies in stark contrast to posterior matching in which the error exponent is infinite for rates below capacity.

%For the parallel AWGN channel with feedback, we could not characterize the second- and third-order asymptotics due to an obstacle to be addressed in the last section of this paper. Instead, we use Chebyshev's inequality
%to prove the strong converse for the parallel AWGN channel with feedback.

\subsection{Paper Outline}
This paper is organized as follows. Section~\ref{notation} summarizes  the notation used in this paper. Section~\ref{sectionDefinition} provides the problem setup of the AWGN channel with feedback under the peak power constraint and presents our main theorem. Section~\ref{sectionPrelim} contains the preliminaries required for the proof of our main theorem, which include important properties of non-asymptotic binary hypothesis testing quantities, and an important lemma concerning {\em simulating output distributions}. Section~\ref{sectionMainResult} presents the proof of our main theorem. Section~\ref{sectionParallelAWGNChannel} discusses the parallel Gaussian channels with feedback under a peak power constraint and applies the techniques used in Section~\ref{sectionMainResult} to prove the strong converse.
%We conclude our discussion and suggest avenues for future research in Section~\ref{sec:conclu}.
\section{Notation}\label{notation}
We use $\Pr\{\mathcal{E}\}$ to represent the probability of an
event~$\mathcal{E}$, and we let $\boldsymbol{1}(\mathcal{E})$ be the characteristic function of $\mathcal{E}$. We use a capital letter~$X$ to denote an arbitrary (can be discrete, continuous or other general) random variable with alphabet $\mathcal{X}$, and use the small letter $x$ to denote a realization of~$X$.
%We use $X_{(1, \ldots, N)}$ to denote a random tuple $(X_1, X_2, \ldots, X_N)$, where each element $X_k$ is chosen from some alphabet $\mathcal{X}_k$ for $k=1, 2, \ldots, N$.
We use $X^n$ to denote a random tuple $(X_1, X_2, \ldots, X_n)$, where the components $X_k$ have the same alphabet~$\mathcal{X}$.

The following notations are used for any arbitrary random variables~$X$ and~$Y$ and any mapping $g$ whose domain includes $\mathcal{X}$. We let $p_X$ and $p_{Y|X}$ denote the probability distribution of $X$ and the conditional probability distribution of $Y$ given $X$ respectively.
%We let $p_X(x)$ and $p_{Y|X}(y|x)$ be the evaluations of $p_X$ and $p_{Y|X}$ respectively at $X=x$ and $Y=y$. We let $p_{g(X)}$ denote the probability distribution of $g(X)$ when $X$ is distributed according to $p_X$, and
We let $\Pr_{p_X}\{g(X)\ge\xi\}$ denote $\int_{x\in \mathcal{X}} p_X(x)\mathbf{1}(\{g(x)\ge\xi\})\, \mathrm{d}x$ for any real-valued function~$g$ and any real constant $\xi$. The expectation and the variance of~$g(X)$ are denoted as
$
\E_{p_X}[g(X)]$ and
$
 \Var_{p_X}[g(X)]\triangleq \E_{p_X}[(g(X)-\E_{p_X}[g(X)])^2]$
 respectively.
  We let $p_Xp_{Y|X}$ denote the joint distribution of $(X,Y)$, i.e., $p_Xp_{Y|X}(x,y)=p_X(x)p_{Y|X}(y|x)$ for all $x$ and $y$.
  %If $X$ and $Y$ are independent, their joint distribution is simply $p_X p_Y$.
  We let $\phi_{\mu,\sigma^2}: \mathbb{R}\rightarrow [0,\infty)$ denote the probability density function of a Gaussian random variable whose mean and variance are $\mu$ and $\sigma^2$ respectively such that
  \[
  \phi_{\mu,\sigma^2}(z) = \frac{1}{\sqrt{2\pi \sigma^2}}e^{-\frac{(z-\mu)^2}{2\sigma^2}}.
  \]
  We will take all logarithms to base 2 throughout this paper.

\section{Additive White Gaussian Noise Channel with Feedback} \label{sectionDefinition}
We consider an additive white Gaussian noise (AWGN) channel with feedback that consists of one source and one destination, denoted by $\mathrm{s}$ and $\, \mathrm{d}$ respectively. Node~$\mathrm{s}$ transmits information to node~$\, \mathrm{d}$ in $n$ time slots as follows. Node~$\mathrm{s}$ chooses message
\[
W\in \{1, 2, \ldots, M\}
 \]
 and sends $W$ to node~$\, \mathrm{d}$, where $M=|\mathcal{W}|$. We assume that $W$ is uniformly distributed over $\{1, 2, \ldots, M\}$. Then for each $k\in \{1, 2, \ldots, n\}$, node~$\mathrm{s}$ transmits $X_{k}\in \mathbb{R}$ in time slot~$k$ and node~$\, \mathrm{d}$ receives
 \[
 Y_{k}=X_k+Z_k,
 \]
 where $Z_1, Z_2, \ldots, Z_n$ are $n$ independent copies of the standard Gaussian random variable. We assume that a noiseless feedback link from~$\, \mathrm{d}$ to~$\mathrm{s}$ exists so that $(W, Y^{k-1})$ is available for encoding $X_k$ at node~$\mathrm{s}$ for each $k\in\{1, 2, \ldots, n\}$. Every codeword $X^n$ transmitted by~$\mathrm{s}$ should satisfy $\sum_{k=1}^nX_{k}^2 \le n P$ where $P>0$ denotes permissible power for $X^n$, i.e., a peak power constraint. In other words, $\Pr\{\sum_{k=1}^nX_{k}^2 \le n P\}=1$. After~$n$ time slots, node~$\, \mathrm{d}$ declares~$\hat W$ to be the transmitted~$W$ based on $Y^n$.
\medskip
\begin{Definition} \label{defCode}
An $(n, M, P)$-feedback code consists of the
following:
\begin{enumerate}
\item A message set
\[
\mathcal{W}\triangleq \{1, 2, \ldots, M\}
\]
 at node~$\mathrm{s}$. Message $W$ is uniform on $\mathcal{W}$.

\item An encoding function
\[
\rho_k : \mathcal{W}\times \mathbb{R}^{k-1}\rightarrow \mathbb{R}
 \]
 for each $k\in\{1, 2, \ldots, n\}$, where $\rho_k$ is the encoding function at node~$\mathrm{s}$ for encoding $X_k$ such that
\[
X_k=\rho_k (W, Y^{k-1})
\]
and
\[
\Pr\left\{\sum_{k=1}^nX_{k}^2 \le nP\right\} = 1.
\]
\item A decoding function
\[
\psi :
\mathbb{R}^{n} \rightarrow \mathcal{W},
 \]
where $\psi$ is the decoding function for $W$ at node~$\, \mathrm{d}$ such that
 \[
 \hat W = \psi(Y^{n}).
 \]
\end{enumerate}
\end{Definition}
\medskip
\begin{Definition}\label{defAWGNchannel}
An additive white Gaussian noise (AWGN) channel with feedback is characterized by the probability density distribution $q_{Y|X}$ satisfying
\begin{equation}
q_{Y|X}(y|x) = \phi_{0, 1}(y-x) \label{defChannelInDefinition}
\end{equation}
such that the following holds for any $(n, M, P)$-feedback code: For each $k\in\{1, 2, \ldots, n\}$,
\begin{align}
\Pr\{W=w, X^k=x^k, Y^k=y^k\}
 = \Pr\{W=w, X^k=x^k, Y^{k-1}=y^{k-1}\}\Pr\{Y_k=y_k|X_k=x_k\} \label{memorylessStatement*}
\end{align}
for all $w$, $x^k$ and $y^k$ where
\begin{equation}
\Pr\{Y_k=y_k|X_k=x_k\} = p_{Y_k|X_k}(y_k|x_k) = q_{Y|X}(y_k|x_k). \label{defChannelInDefinition*}
\end{equation}
Since $p_{Y_k|X_k}$ does not depend on~$k$ by \eqref{defChannelInDefinition*}, the channel is stationary.
\end{Definition}
\medskip

 For any $(n, M, P)$-feedback code defined on the AWGN channel with feedback, let $p_{W,X^n, Y^n, \hat W}$ be the joint distribution induced by the code. We can factorize $p_{W,X^n, Y^n, \hat W}$ as follows:
\begin{align}
 p_{W,X^n, Y^n, \hat W}
&\stackrel{\text{(a)}}{=} p_{W,X^n, Y^n}p_{\hat W |Y^n}  \notag\\
&= p_W \left(\prod_{k=1}^np_{X_k,Y_k|X^{k-1}, Y^{k-1}, W}\right)p_{\hat W |Y^n} \notag\\
&= p_W \left(\prod_{k=1}^n p_{X_k|X^{k-1}, Y^{k-1}, W}p_{Y_k|X^k, Y^{k-1}, W} \right)p_{\hat W |Y^n} \notag\\
& \stackrel{\text{(b)}}{=} p_W \left(\prod_{k=1}^n \left( p_{X_k|W, Y^{k-1}} p_{Y_k|X^k, Y^{k-1}, W} \right)\right)p_{\hat W |Y^n}\notag\\
& \stackrel{\text{(c)}}{=} p_W \left(\prod_{k=1}^n \left(p_{X_k|W, Y^{k-1}} p_{Y_k|X_k}\right)\right)p_{\hat W |Y^n}. \label{memorylessStatement}
\end{align}
where
\begin{enumerate}
\item[(a)] follows from Definition~\ref{defCode} that $\hat W$ is a function of $Y^n$.
\item[(b)] follows from Definition~\ref{defCode} that $X_k$ is a function of $(W, Y^{k-1})$ for each $k\in\{1, 2, \ldots, n\}$.
   \item[(c)] follows from \eqref{memorylessStatement*} and \eqref{defChannelInDefinition*} that for all $w$, $x^k$ and $y^k$ such that $p_{X^k, Y^{k-1}, W}(x^k, y^{k-1}, w)>0$,
   \begin{equation}
    p_{Y_k|W, X^k, Y^{k-1}}(y_k|w, x^k, y^{k-1}) = p_{Y_k|X_k}(y_k|x_k) = q_{Y|X}(y_k|x_k). \label{memorylessStatement+}
    \end{equation}
\end{enumerate}
 % It can be derived from Definition~\ref{defAWGNchannel} that $\{Y_k-X_k\}_{k=1}^n$ are independent standard Gaussian random variables, which is shown below for completeness. For each $k\in\{1, 2, \ldots, n\}$,
% \begin{align*}
%& p_{W, X^k, Y^{k-1}, Y_k-X_k}(w, x^k, y^{k-1}, y_k-x_k) \notag\\
%& = p_{W, X^k, Y^{k-1}}(w, x^k, y^{k-1}) p_{Y_k-X_k|W, X^k, Y^{k-1}}(y_k-x_k|w, x^k, y^{k-1}) \notag\\
%& = p_{W, X^k, Y^{k-1}}(w, x^k, y^{k-1}) p_{Y_k|W, X^k, Y^{k-1}}(y_k|w, x^k, y^{k-1}) \notag\\
%&\stackrel{\eqref{memorylessStatement*}}{=}p_{W, X^k, Y^{k-1}}(w, x^k, y^{k-1})q_{Y_k|X_k}(y_k|x_k)\notag\\
%&  \stackrel{\eqref{defChannelInDefinition}}{=}p_{W, X^k, Y^{k-1}}(w, x^k, y^{k-1})\phi_{(Y_k-X_k; 0,1)}(y_k-x_k)
% \end{align*}
% for all $w$, $x^k$ and $y^k$, which implies that
% \begin{equation}
% p_{Y_k-X_k|W, X^k, Y^{k-1}} = p_{Y_k-X_k} = \phi_{(Y_k-X_k; 0,1)}. \label{distributionOfYk-Xk}
% \end{equation}
% Consequently,
% \begin{align}
% p_{\{Y_k-X_k: k\in \{1, 2, \ldots, n\} \}}
%& = \prod_{k=1}^n p_{Y_k-X_k|\{Y_\ell-X_\ell\}_{\ell=1}^{k-1}} \notag\\
%& \stackrel{\eqref{distributionOfYk-Xk}}{=} \prod_{k=1}^n \phi_{(Y_k-X_k; 0,1)}. \label{independentNoises}
% \end{align}
\medskip
\begin{Definition} \label{defErrorProbability}
For an $(n, M, P)$-feedback code defined on the AWGN channel with feedback, we can calculate according to \eqref{memorylessStatement} the \textit{average probability of decoding error} defined as $\Pr\left\{\hat W \ne W\right\}$.
We call an $(n, M, P)$-feedback code with average probability of decoding error no larger than $\varepsilon$ an $(n, M, P, \varepsilon)$-feedback code.
\end{Definition}
\medskip

Before stating our main result, we define $\Phi: (-\infty, \infty)\rightarrow (0,1)$ to be the cumulative distribution function for the standard Gaussian distribution and recall the definitions of $\mathrm{C}(P)$ and $\mathrm{V}(P)$
in \eqref{defCP} and \eqref{defVP}. Since $\Phi$ is strictly increasing on $(-\infty, \infty)$, the inverse of $\Phi$ is well-defined and is denoted by $\Phi^{-1}$.
% is defined as
% \begin{equation}
% \Phi^{-1}(\varepsilon) \triangleq \sup\{y\in \mathbb{R}: \Phi(y)\le \varepsilon\}. \label{defPhiInverse}
% \end{equation}
The following theorem is the main result in this paper.
\medskip
\begin{Theorem} \label{thmMainResult}
Fix an $\varepsilon \in (0,1)$ and let
\[
M_{\mathrm{fb}}^*(n, \varepsilon, P) \triangleq \max\{M: \text{There exists an $(n, M, P, \varepsilon)$-feedback code}\}.
\]
Then, there exists a constant $\kappa$ not depending on~$n$ such that for each $n\in \mathbb{N}$,
\begin{equation}
\log M_{\mathrm{fb}}^*(n, \varepsilon, P) \le n \mathrm{C}(P) + \sqrt{n\mathrm{V}(P)}\Phi^{-1}(\varepsilon) +  \frac{1}{2}\log n + \kappa. \label{cutsetStatement}
\end{equation}
\end{Theorem}
\medskip
Combining \eqref{eqn:asymp_expans} and Theorem~\ref{thmMainResult}, we complete the characterizations of the first-, second- and third-order asymptotics for the AWGN channel with feedback as shown in \eqref{eqn:asymp_expans_fb}.

In order to prove our main theorem, we need to leverage important properties of the non-asymptotic quantities in binary hypothesis testing and we also need to construct so-called {\em simulating output distributions}. These preliminaries   are contained in Section~\ref{sectionPrelim}. The details of the proof of Theorem~\ref{thmMainResult} are provided in Section~\ref{sectionMainResult}.

\section{Preliminaries for the Proof of Theorem \ref{thmMainResult}} \label{sectionPrelim}
\subsection{Binary Hypothesis Testing} \label{sectionBHT}
The following definition concerning the non-asymptotic fundamental limits of a simple binary hypothesis test  is standard. See for example  \cite[Section~2.3]{Pol10}.
\medskip
\begin{Definition}\label{defBHTDivergence}
Let $p_{X}$ and $q_{X}$ be two probability distributions on some common alphabet $\mathcal{X}$. Let
\[
\mathcal{A}(\{0,1\}|\mathcal{X})\triangleq \{
r_{Z|X}: \text{$Z$ and $X$ assume values in $\{0,1\}$ and $\mathcal{X}$ respectively}\}
\]
be the set of randomized binary hypothesis tests between $p_{X}$ and $q_{X}$ where $\{Z=0\}$ indicates the test chooses $q_X$, and let $\delta\in [0,1]$ be a real number. The minimum type-II error in a simple binary hypothesis test between $p_{X}$ and $q_{X}$ with type-I error no larger than $1-\delta$ is defined as
\begin{align}
 \beta_{\delta}(p_X\|q_X) \triangleq
\inf\limits_{\substack{r_{Z|X} \in \mathcal{A}(\{0,1\}|\mathcal{X}): \\ \int_{x\in \mathcal{X}}r_{Z|X}(1|x)p_X(x)\, \mathrm{d}x\ge \delta}} \int_{x\in \mathcal{X}}r_{Z|X}(1|x)q_X(x)\, \mathrm{d}x.\label{eqDefISDivergence}
\end{align}
\end{Definition}\medskip
The existence of a minimizing test $r_{Z|X}$ is guaranteed by the Neyman-Pearson lemma.

We state in the following lemma and proposition some important properties of $\beta_{\delta}(p_X\|q_X)$, which are crucial for the proof of Theorem~\ref{thmMainResult}. The proof of the following lemma can be found  in, for example, Wang-Colbeck-Renner~\cite[Lemma~1]{Wang2009}.
\medskip
\begin{Lemma}\label{lemmaDPI} Let $p_{X}$ and $q_{X}$ be two probability distributions on some $\mathcal{X}$, and let $g$ be a function whose domain contains $\mathcal{X}$. Then, the following two statements hold:
\begin{enumerate}
\item[1.] (Data processing inequality (DPI)) $\beta_{\delta}(p_X\|q_X) \le \beta_{\delta}(p_{g(X)}\|q_{g(X)})$.
\item[2.] For all $\xi>0$, $\beta_{\delta}(p_X\|q_X)\ge \frac{1}{\xi}\left(\delta - \int_{x\in\mathcal{X}}p_X(x) \boldsymbol{1}\left(\left\{ \frac{p_X(x)}{q_X(x)} \ge \xi \right\}\right)\, \mathrm{d}x\right) $.
\end{enumerate}
\end{Lemma}
\medskip
 The proof of the following   proposition can be also be  found in Wang-Colbeck-Renner~\cite[Lemma 3]{Wang2009}.
 \medskip
\begin{Proposition} \label{propositionBHTLowerBound}
Let $p_{U,V}$ be a probability distribution defined on $\mathcal{W}\times \mathcal{W}$ for some finite alphabet $\mathcal{W}$, and let $p_U$ be the marginal distribution of $p_{U,V}$. In addition, let $q_{V}$ be a distribution defined on $\mathcal{W}$. Suppose $p_{U}$ is the uniform distribution, and let
\begin{equation}
\alpha = \Pr\{U\ne V\} \label{defAlpha}
\end{equation}
be a real number in $[0, 1)$ where $(U,V)$ is distributed according to $p_{U,V}$. Then,
\begin{equation}
|\mathcal{W}| \le 1/\beta_{1-\alpha}(p_{U,V}\|p_{U} q_{V}). \label{propositionBHTLowerBoundEq1}
\end{equation}
\end{Proposition}

\subsection{Simulating Output Distribution} \label{sectionSimulatingDistribution}
Proposition~\ref{propositionBHTLowerBound} and Statement~2 of Lemma~\ref{lemmaDPI} together imply a lower bound for the error probability, and the lower bound holds for all $q_V$. Therefore, we are motivated to choose a \textit{simulating output distribution} $q_V$ which is almost the same as the output distribution chosen in \cite[Section 4.2.2]{Pol10} so that the right hand side of \eqref{propositionBHTLowerBoundEq1} can be simplified. The construction of the simulating output distribution is contained in the following lemma.

\begin{Lemma} \label{lemmaSimulatingDistribution}
 Given an $(n, M, P, \varepsilon)$-feedback code for the AWGN channel, let $p_{W,X^n, Y^n, \hat W}$ be the probability distribution induced by the code according to \eqref{memorylessStatement}. Then there exists a probability distribution $s_{Y^n, \hat W}$ that satisfies the following properties:
  \begin{enumerate}
 %\item[(i)] $s_{W} = p_{W}$
 \item[(i)] $s_{\hat W|Y^n} = p_{\hat W|Y^n}$
 \item [(ii)] $s_{Y^n} = \prod_{k=1}^n s_{Y_k}$
     \item[(iii)] For each $k\in\{1, 2, \ldots, n\}$, $s_{Y_k}(y_k) = \phi_{0, 1+P}(y_k)$ for all $y_k\in \mathbb{R}$.
 \end{enumerate}
We call $s_{Y^n, \hat W}$ a \textit{simulating output distribution of $p_{W,X^n, Y^n, \hat W}$} because $s_{Y^n, \hat W}$ captures all the important properties of $(Y^n, \hat W)$ when $(W, X^n, Y^n, \hat W)$ is generated according to the given probability distribution $p_{W,X^n, Y^n, \hat W}$.
\end{Lemma}
\begin{IEEEproof}
Define $s_{Y^n, \hat W}$ as
  \begin{equation}
 s_{Y^n, \hat W}(y^n, \hat w)= \left( \prod_{k=1}^n \phi_{0, 1+P}(y_k) \right) p_{\hat W|Y^n}(\hat w|y^n) \label{marginal2ndWay}
 \end{equation}
 for all $\hat W \in \mathcal{W}$ and $y^n\in \mathbb{R}^n$.
  In order to prove Property~(i), we marginalize \eqref{marginal2ndWay} and obtain
   \begin{equation}
 s_{Y^n}(y^n)= \prod_{k=1}^n \phi_{0, 1+P}(y_k)  \label{marginal3rdWay}
 \end{equation}
 for all $y^n\in \mathbb{R}^n$. Property~(i) then follows from \eqref{marginal2ndWay} and \eqref{marginal3rdWay}. Property (iii) follows from marginalizing \eqref{marginal3rdWay}. Property (ii) follows from \eqref{marginal3rdWay} and  Property (iii).
\end{IEEEproof}

\section{Proof of Theorem~\ref{thmMainResult}} \label{sectionMainResult}
\subsection{Lower Bounding the Error Probability in Terms of the Type-II Error  of a Hypothesis Test}
Fix an $\varepsilon\in (0,1)$ and choose an arbitrary sequence of $(\bar n, M_{\mathrm{fb}}^*(\bar n, \varepsilon, P), P, \varepsilon)$-feedback codes for the AWGN channel with feedback. Using Definition~\ref{defCode}, we have
\begin{equation}
\Pr\left\{\sum_{k=1}^{\bar n}X_{k}^2 \le \bar nP\right\} = 1 \label{powerConstraintInProof}
\end{equation}
for the $(\bar n, M_{\mathrm{fb}}^*(\bar n, \varepsilon, P), P, \varepsilon)$-feedback code for each~$\bar n\in \mathbb{N}$. Given the  $(\bar n, M_{\mathrm{fb}}^*(\bar n, \varepsilon, P), P, \varepsilon)$-feedback code, we can always construct an  $(\bar n + 1, M_{\mathrm{fb}}^*(\bar n, \varepsilon, P), P, \varepsilon)$-feedback code by appending a carefully chosen $X_{\bar n+1}$ to each transmitted codeword $X^{\bar n}$ generated by the $(\bar n, M_{\mathrm{fb}}^*(\bar n, \varepsilon, P), P, \varepsilon)$-feedback code such that
\begin{equation}
\Pr\left\{\sum_{k=1}^{\bar n+1}X_{k}^2 = (\bar n+1)P\right\} = 1. \label{powerConstraintInProofnBar}
\end{equation}
The technique of transforming the peak power inequality constraint \eqref{powerConstraintInProof} to a power equality constraint \eqref{powerConstraintInProofnBar} by appending an extra symbol has been employed in \cite[Lemma~39]{PPV10} and \cite[Theorem~4.4]{TanBook} (and is called the Yaglom map trick).
To simplify notation, we let $n=\bar n+1$.
 Let $p_{W,X^n, Y^n, \hat W}$ be the probability distribution induced by the $(n, M_{\mathrm{fb}}^*(n-1, \varepsilon, P), P, \varepsilon)$-feedback code constructed above for each $n\in\{2, 3, \ldots\}$, where $p_{W,X^n, Y^n, \hat W}$ is obtained according to \eqref{memorylessStatement}. In view of \eqref{powerConstraintInProofnBar}, we assume without loss of generality that
 \begin{equation}
p_{W,X^n, Y^n}(w,x^n, y^n) = p_{W,X^n, Y^n}(w,x^n, y^n)\mathbf{1}\left(\left\{\sum_{k=1}^{n}x_k^2 = nP\right\}\right) \label{powerConstraintCharacteristicFunction}
 \end{equation}
 for all $w\in\mathcal{W}$, $x^n\in \mathbb{R}^n$ and $y^n\in\mathbb{R}^n$.
  Fix an $(n, M_{\mathrm{fb}}^*(n-1, \varepsilon, P), P, \varepsilon)$-feedback code.
Let $s_{Y^n, \hat W}$ be a simulating output distribution of $p_{W,X^n, Y^n, \hat W}$ such that $s_{Y^n, \hat W}$ satisfies all the properties in Lemma~\ref{lemmaSimulatingDistribution}. Then, it follows from Proposition~\ref{propositionBHTLowerBound} and Definition~\ref{defCode} with the identifications $U\equiv W$, $V\equiv \hat W$, $p_{U,V}\equiv p_{W,\hat W}$, $q_V\equiv s_{\hat W}$, $|\mathcal{W}|\equiv M_{\mathrm{fb}}^*(n-1, \varepsilon, P)$ and $\alpha\equiv \Pr\{\hat W \ne W\} \le \varepsilon$ that
 \begin{align}
\beta_{1-\varepsilon}(p_{W,\hat W}\|p_W s_{\hat W}) \le \beta_{1-\alpha}(p_{W,\hat W}\|p_W s_{\hat W}) \le 1/M_{\mathrm{fb}}^*(n-1, \varepsilon, P). \label{eqnBHTReverseChain}
 \end{align}
  \subsection{Using the DPI to Introduce the Channel Input and Output}
Consider the following chain of inequalities:
 \begin{align}
 & \beta_{1-\varepsilon}(p_{W,\hat W}\|p_W s_{\hat W}) \notag\\
& =\beta_{1-\varepsilon}(p_W p_{\hat W|W}\|p_W s_{\hat W}) \notag\\
&\stackrel{\text{(a)}}{\ge} \beta_{1-\varepsilon}(p_Wp_{\hat W, Y^n|W}\|p_W s_{\hat W, Y^n}) \notag\\
& = \beta_{1-\varepsilon}(p_Wp_{Y^n|W}p_{\hat W|Y^n, W}\|p_Ws_{Y^n} s_{\hat W|Y^n}) \notag\\
& \stackrel{\text{(b)}}{=} \beta_{1-\varepsilon}(p_W p_{Y^n|W}p_{\hat W|Y^n, W}\|p_W s_{Y^n} p_{\hat W|Y^n}) \notag\\
& \stackrel{\text{(c)}}{=}\beta_{1-\varepsilon}(p_W p_{Y^n|W}p_{\hat W|Y^n}\|p_W s_{Y^n} p_{\hat W|Y^n}) \notag\\
&\stackrel{\text{(d)}}{\ge} \beta_{1-\varepsilon}\left(p_W p_{\hat W|Y^n}p_{X^n,Y^n|W}\left\|p_Wp_{\hat W|Y^n} s_{Y^n}\prod_{k=1}^n p_{X_k|Y^{k-1}, W}\right.\right)  \label{eqnBHTFirstChain}
\end{align}
where
 \begin{enumerate}
 \item[(a)] follows from the DPI of $\beta_{1-\varepsilon}$ by introducing the channel output $Y^n$.
 \item[(b)] follows from Property~(i) in Lemma~\ref{lemmaSimulatingDistribution}.
 \item[(c)] follows from the fact that
 \[
 W  \rightarrow Y^n \rightarrow \hat W
 \]
 forms a Markov chain for the $(n, M_{\mathrm{fb}}^*(n-1, \varepsilon, P), P, \varepsilon)$-feedback code (cf.\ Definition~\ref{defCode}).
 \item[(d)] follows from the DPI of $\beta_{1-\varepsilon}$ by introducing the channel input $X^n$.
 \end{enumerate}
  \subsection{Obtaining a Non-Asymptotic Bound from the Binary Hypothesis Testing}
 Following \eqref{eqnBHTFirstChain}, we marginalize \eqref{memorylessStatement} and obtain
 \[
   p_{W, X^n , Y^n} =p_W \prod_{k=1}^n (p_{X_k|Y^{k-1}, W}p_{Y_k|X_k})
 \]
 which implies that
\begin{align}
  p_{X^n , Y^n|W} =\prod_{k=1}^n (p_{X_k|Y^{k-1}, W}p_{Y_k|X_k}).  \label{eqnBHTFirstChain*}
 \end{align}
 Combining \eqref{eqnBHTFirstChain} and \eqref{eqnBHTFirstChain*}, we have
 \begin{align}
&  \beta_{1-\varepsilon}(p_{W,\hat W}\|p_W s_{\hat W}) \notag\\
& \ge \beta_{1-\varepsilon}\left(p_W p_{\hat W|Y^n}\prod_{k=1}^n (p_{X_k|Y^{k-1}, W}p_{Y_k|X_k})\left\|p_{\hat W|Y^n} p_W s_{Y^n} \prod_{k=1}^n ( p_{X_k|Y^{k-1}, W} )\right.\right) \notag\\
& \stackrel{\text{(a)}}{=} \beta_{1-\varepsilon}\left(p_W p_{\hat W|Y^n}\prod_{k=1}^n (p_{X_k|Y^{k-1}, W}p_{Y_k|X_k})\left\|p_W p_{\hat W|Y^n} \prod_{k=1}^n ( p_{X_k|Y^{k-1}, W} s_{Y_k})\right.\right) \label{eqnBHT5thChain}
 \end{align}
 where (a) follows from Property~(ii) in Lemma~\ref{lemmaSimulatingDistribution}.
Fix any constant $\xi_n>0$ to be specified later. Using Lemma~\ref{lemmaDPI} and \eqref{eqnBHTFirstChain*},
we have
\begin{align}
&\beta_{1-\varepsilon}\left(p_W p_{\hat W|Y^n}\prod_{k=1}^n (p_{X_k|Y^{k-1}, W}p_{Y_k|X_k})\left\|p_W p_{\hat W|Y^n} \prod_{k=1}^n ( p_{X_k|Y^{k-1}, W} s_{Y_k})\right.\right)  \notag\\
&\ge \frac{1}{\xi_n}\left(1-\varepsilon - \int_{w, x^n, y^n}p_{W, X^n, Y^n}(w, x^n, y^n) \boldsymbol{1}\left(\left\{\prod_{k=1}^n \frac{p_{Y_k|X_k}(y_k|x_{k})}{s_{Y_{k}}(y_k)} \ge \xi_n \right\}\right)\, \mathrm{d}y^n \, \mathrm{d}x^n \, \mathrm{d}w\right). \label{eqnBHTThirdChain}
\end{align}
Combining \eqref{eqnBHTReverseChain}, \eqref{eqnBHT5thChain} and \eqref{eqnBHTThirdChain}, we have
\begin{align}
&\log M_{\mathrm{fb}}^*(n-1, \varepsilon, P)\notag\\
& \le \log\xi_n - \log\left(1-\varepsilon - \int_{w,x^n, y^n}p_{W,X^n, Y^n}(w,x^n, y^n) \boldsymbol{1}\left(\left\{\sum_{k=1}^n \log\frac{p_{Y_k|X_k}(y_k|x_{k})}{s_{Y_{k}}(y_k)} \ge \log\xi_n \right\}\right)\, \mathrm{d}y^n \, \mathrm{d}x^n \, \mathrm{d}w\right) \notag\\
& = \log\xi_n - \log\left(1-\varepsilon - \Pr_{p_{W, X^n, Y^n}} \left\{\sum_{k=1}^n \log\frac{p_{Y_k|X_k}(Y_{k}|X_{k})}{s_{Y_{k}}(Y_{k})} \ge \log\xi_n \right\}\right) \notag\\
& = \log\xi_n - \log\left(\Pr_{p_{W, X^n, Y^n}} \left\{\sum_{k=1}^n \log\frac{p_{Y_k|X_k}(Y_{k}|X_{k})}{s_{Y_{k}}(Y_{k})} < \log\xi_n \right\}-\varepsilon\right).\label{eqnBHTSecondChain}
\end{align}
\subsection{Simplifying the Non-Asymptotic Bound}
The channel law is
\begin{equation}
p_{Y_k|X_k}(y_{k}|x_{k}) = \phi_{0,1}(y_k-x_k) \label{memorylessStatement++}
\end{equation}
for each $k\in\{1, 2, \ldots, n\}$.
Combining \eqref{memorylessStatement++} and Property~(iii) in Lemma~\ref{lemmaSimulatingDistribution}, we have
\begin{equation}
\log\frac{p_{Y_k|X_k}(Y_{k}|X_{k})}{s_{Y_{k}}(Y_{k})} = \frac{1}{2}\log(1+P)+ \frac{\log e}{2(1+P)}\left(-P(Y_k-X_k)^2 + X_k^2 + 2X_k(Y_k-X_k)\right) \label{defLogLikelihood}
\end{equation}
for each $k\in\{1, 2, \ldots, n\}$. Due to the power equality constraint imposed on the codewords, we have
\begin{equation}
\Pr_{p_{W, X^n, Y^n}}\left\{ \sum_{k=1}^n  X_k^2 = nP \right\} \stackrel{\eqref{powerConstraintInProofnBar}}{=}1. \label{powerConstraintInProofn}
\end{equation}
 Letting
 \begin{equation}
U_k \triangleq \frac{\log e}{2(1+P)}(-P(Y_k-X_k)^2 + 2X_k(Y_k-X_k)+P) \label{defUk}
 \end{equation}
 for each $k\in\{1, 2, \ldots, n\}$, it follows from \eqref{defLogLikelihood} and \eqref{powerConstraintInProofn} that
\begin{equation}
\Pr_{p_{W, X^n, Y^n}}\left\{\sum_{k=1}^n \log\frac{p_{Y_k|X_k}(Y_{k}|X_{k})}{s_{Y_{k}}(Y_{k})} =\frac{n}{2}\log(1+P) + \sum_{k=1}^n U_k\right\}=1.\label{expectationSumUk}
\end{equation}
Combining \eqref{eqnBHTSecondChain} and \eqref{expectationSumUk}, we have
\begin{align}
\log M_{\mathrm{fb}}^*(n-1, \varepsilon, P) \le \log\xi_n - \log\left(\Pr_{p_{W, X^n, Y^n}} \left\{\sum_{k=1}^n U_k < \log\xi_n - \frac{n}{2}\log(1+P) \right\}-\varepsilon\right). \label{eqnBHT6thChain}
\end{align}
%It can be shown from \eqref{memorylessStatement*} and \eqref{defUk} that
%\[
%\E_{p_{W,X^n, Y^n}}\left[ U_k |X^{k-1}, Y^{k-1}\right] =0
%\]
%for each $k\in\{1, 2, \ldots, n\}$, which implies that $\sum_{k=1}^n U_k$ is a martingale.
\subsection{Evaluating the Distribution of the Sum of Random Variables $\sum_{k=1}^n U_k$}
In order to simplify \eqref{eqnBHT6thChain}, we now investigate the distribution of the sum of random variables $\sum_{k=1}^n U_k$. Note that if the AWGN channel has no feedback, it follows from spherical symmetry \cite[Section 4.2.2]{Pol10} of the AWGN channel   that the evaluation of $\eqref{eqnBHT6thChain}$ can be simplified by assuming without loss of generality that
\begin{equation}
(X_1, X_2, \ldots, X_n) = (\sqrt{P}, \sqrt{P}, \ldots, \sqrt{P}). \label{powerAssumption}
 \end{equation}
Surprisingly in the feedback case, we will show in the following that the distribution of $\sum_{k=1}^n U_k$ can be evaluated in closed form.  We need not  appeal to any sophisticated Berry-Ess\'een-type results for bounded martingale difference sequences~\cite{Machkouri} as was done by Altu\u{g} and Wagner for discrete memoryless channels in \cite{AW14}.
% not affected by imposing the extra assumption \eqref{powerAssumption} and
The evaluation of $\eqref{eqnBHT6thChain}$ is as simple as the no-feedback case. Define the function
$
\lambda: \mathbb{R}\times \mathbb{R}\rightarrow \mathbb{R}
$
\begin{equation}
\lambda(x, y)= -P(y-x)^2 + 2 x(y-x). \label{defLambda}
\end{equation}
We begin evaluating the distribution of $\sum_{k=1}^n U_k$ by examining the distribution of
$\sum_{k=1}^n \lambda(X_k, Y_k)$ (cf.\ \eqref{defUk}) as follows.
Let
\begin{equation}
\E_{p_{W, X^n, Y^n}}\left[e^{t\sum_{k=1}^n \lambda(X_k, Y_k)}\right] \label{momentGenPartialUk}
\end{equation}
 be the moment generating function of $\sum_{k=1}^n \lambda(X_k, Y_k)$. In order to evaluate a closed form expression for \eqref{momentGenPartialUk}, we write
 \begin{align}
 &\E_{p_{W, X^n, Y^n}}\left[e^{t\sum_{k=1}^n \lambda(X_k, Y_k)}\right]  \notag\\
&= \int_{w, x^{n}, y^{n}} p_{W, X^{n}, Y^{n}}(w, x^{n},y^{n})e^{t\sum_{k=1}^{n} \lambda(x_k,y_k)}
  \, \mathrm{d}y^n \, \mathrm{d}x^n\, \mathrm{d}w \notag\\
  & \stackrel{\eqref{powerConstraintCharacteristicFunction}}{=} \int_{w, x^{n}, y^{n}} p_{W, X^{n}, Y^{n}}(w, x^{n},y^{n}) \mathbf{1}\left(\left\{\sum_{k=1}^{n}x_k^2 = nP\right\}\right)  e^{t\sum_{k=1}^{n} \lambda(x_k,y_k)}
  \, \mathrm{d}y^n \, \mathrm{d}x^n\, \mathrm{d}w \notag\\
  & = \int\limits_{w, x^{n}, y^{n}} p_{W, X^{n}, Y^{n}}(w, x^{n},y^{n}) \mathbf{1}\left(\left\{\sum_{k=1}^{n}x_k^2 = nP\right\}\right)  e^{t\left(\sum\limits_{k=1}^{n} \lambda(x_k,y_k)\right)+\frac{2t^2}{1+2tP} \left(nP - \sum\limits_{k=1}^{n}x_k^2\right)}
  \, \mathrm{d}y^n \, \mathrm{d}x^n\, \mathrm{d}w \notag\\
   & \stackrel{\eqref{powerConstraintCharacteristicFunction}}{=} \int\limits_{w, x^{n}, y^{n}} p_{W, X^{n}, Y^{n}}(w, x^{n},y^{n}) e^{t\left(\sum\limits_{k=1}^{n} \lambda(x_k,y_k)\right)+\frac{2t^2}{1+2tP} \left(nP - \sum\limits_{k=1}^{n}x_k^2\right)}
  \, \mathrm{d}y^n \, \mathrm{d}x^n\, \mathrm{d}w \notag\\
  &  \stackrel{\text{(a)}}{=}\frac{1}{|\mathcal{W}|}\sum\limits_{w\in \mathcal{W}}\int_{x^{n}, y^{n}} p_{X^{n}, Y^{n}|W}(x^{n},y^{n}|w)e^{t\left(\sum\limits_{k=1}^{n} \lambda(x_k,y_k)\right)+\frac{2t^2}{1+2tP} \left(nP - \sum\limits_{k=1}^{n}x_k^2\right)}
  \, \mathrm{d}y^n \, \mathrm{d}x^n \label{eqnBHT4thChain}
\end{align}
 where (a) follows from Definition~\ref{defCode} that $W$ is uniform on $\mathcal{W}$.
% \begin{enumerate}
% \item[(a)] follows from \eqref{powerConstraintInProofn} that
%%\[
%%\Pr_{p_{W,X^n, Y^n}}\left\{\sum_{k=1}^n  X_k^2 = nP \right\}=1.
%%\]
%\begin{equation}
%\frac{2t^2}{1+2tP} \left(nP - \sum\limits_{k=1}^{n}x_k^2\right)=0 \label{stepAeqnBHT4thChain}
%\end{equation}
%for all $x^n$ on the support of $p_{X^n}$.
 Following \eqref{eqnBHT4thChain}, consider the following chain of equalities for each $w\in \mathcal{W}$ and each $\ell\in \{0, 1, \ldots, n-2\}$:
 \begin{align}
 & \int\limits_{x^{n-\ell}, y^{n-\ell}} p(x^{n-\ell},y^{n-\ell}|w)e^{t\left(\sum\limits_{k=1}^{n-\ell} \lambda(x_k,y_k)\right) +\frac{2t^2}{1+2tP} \left(nP - \sum\limits_{k=1}^{n-\ell}x_k^2\right)}
  \, \mathrm{d}y^{n-\ell} \, \mathrm{d}x^{n-\ell} \notag\\
 & = \int\limits_{x^{n-\ell-1}, y^{n-\ell-1}} p(x^{n-\ell-1},y^{n-\ell-1}|w)e^{t\left(\sum\limits_{k=1}^{n-\ell-1} \lambda(x_k,y_k)\right) +\frac{2t^2}{1+2tP} \left(nP - \sum\limits_{k=1}^{n-\ell-1}x_k^2\right)} \notag\\
 & \qquad \times
   \int\limits_{x_{n-\ell}, y_{n-\ell}}   p(x_{n-\ell},y_{n-\ell}|w, x^{n-\ell-1}, y^{n-\ell-1}) e^{t\lambda(x_{n-\ell},y_{n-\ell}) -\frac{2t^2x_{n-\ell}^2}{1+2tP} }\, \mathrm{d}y_{n-\ell} \, \mathrm{d}x_{n-\ell} \, \mathrm{d}y^{n-\ell-1} \, \mathrm{d}x^{n-\ell-1}\notag\\
   & \stackrel{\text{(a)}}{=} \int\limits_{x^{n-\ell-1}, y^{n-\ell-1}} p(x^{n-\ell-1},y^{n-\ell-1}|w)e^{t\left(\sum\limits_{k=1}^{n-\ell-1} \lambda(x_k,y_k)\right) +\frac{2t^2}{1+2tP} \left(nP - \sum\limits_{k=1}^{n-\ell-1}x_k^2\right)} \notag\\
 & \qquad \times
   \int\limits_{x_{n-\ell}}   p(x_{n-\ell}|w,x^{n-\ell-1}, y^{n-\ell-1}) e^{\frac{-2t^2 x_{n-\ell}^2}{1+2tP}} \notag\\
    &\qquad \times \int\limits_{y_{n-\ell}} \phi_{0,1}(y_{n-\ell}-x_{n-\ell}) e^{t \lambda(x_{n-\ell},y_{n-\ell})}\, \mathrm{d}y_{n-\ell} \, \mathrm{d}x_{n-\ell} \, \mathrm{d}y^{n-\ell-1} \, \mathrm{d}x^{n-\ell-1}\notag\\
   & \stackrel{\text{(b)}}{=} \int\limits_{x^{n-\ell-1}, y^{n-\ell-1}} p(x^{n-\ell-1},y^{n-\ell-1}|w)e^{t\left(\sum\limits_{k=1}^{n-\ell-1} \lambda(x_k,y_k)\right) +\frac{2t^2}{1+2tP} \left(nP - \sum\limits_{k=1}^{n-\ell-1}x_k^2\right)} \notag\\
 & \qquad \times
   \int\limits_{x_{n-\ell}}   p(x_{n-\ell}|w,x^{n-\ell-1}, y^{n-\ell-1}) e^{\frac{-2t^2 x_{n-\ell}^2}{1+2tP}} \notag\\
    &\qquad \times \frac{1}{\sqrt{1+2tP}}e^{\frac{2t^2x_{n-\ell}^2}{1+2tP}} \, \mathrm{d}x_{n-\ell} \, \mathrm{d}y^{n-\ell-1} \, \mathrm{d}x^{n-\ell-1}\notag\\
   &   = \frac{1}{\sqrt{1+2tP}}\int\limits_{x^{n-\ell-1}, y^{n-\ell-1}} p(x^{n-\ell-1},y^{n-\ell-1}|w)e^{t\left(\sum\limits_{k=1}^{n-\ell-1} \lambda(x_k,y_k)\right) +\frac{2t^2}{1+2tP} \left(nP - \sum\limits_{k=1}^{n-\ell-1}x_k^2\right)} \, \mathrm{d}y^{n-\ell-1} \, \mathrm{d}x^{n-\ell-1} \label{reverseInductionStep1}
 \end{align}
 where
 \begin{enumerate}
 \item[(a)] follows from \eqref{memorylessStatement*} and \eqref{memorylessStatement++}.
 \item[(b)] follows from evaluating the integral
\begin{align*}
& \int\limits_{y_{n-\ell}} \phi_{0,1}(y_{n-\ell}-x_{n-\ell}) e^{t \lambda(x_{n-\ell},y_{n-\ell})}\, \mathrm{d}y_{n-\ell} \\*
 & = \frac{1}{\sqrt{2\pi}} \int\limits_{y_{n-\ell}}   e^{- (y_{n-\ell}-x_{n-\ell})^2/2}e^{t(-P(y_{n-\ell}-x_{n-\ell})^2+2 x_{n-\ell}(y_{n-\ell}-x_{n-\ell}) ) }\,   \mathrm{d}y_{n-\ell} \\
 &=\int\limits_{z}\frac{1}{\sqrt{2\pi}} e^{-z^2/2}e^{t ( -Pz^2 + 2 x_{n-\ell} z)}\, \mathrm{d}z \\
 & =\frac{1}{\sqrt{2\pi}} \sqrt{ \frac{\pi}{\frac{1}{2} + tP}}e^{ \frac{(2tx_{n-\ell})^2 }{4 \left(\frac{1}{2}+tP\right)}} \\
&= \frac{1}{\sqrt{1+2tP}}e^{\frac{2t^2x_{n-\ell}^2}{1+2tP}}
\end{align*}
 by using the definition of $\lambda(\cdot,\cdot)$ in  \eqref{defLambda} and the substitution
 \[
 \phi_{0,1}(y_{n-\ell}-x_{n-\ell})=\frac{1}{\sqrt{2\pi}}e^{-(y_{n-\ell}-x_{n-\ell})^2/2}.
 \]
 \end{enumerate}
Applying \eqref{reverseInductionStep1} recursively from $\ell=0$ to $\ell=n-2$, we have for each $w\in\mathcal{W}$
\begin{align}
& \int_{x^{n}, y^{n}} p_{X^{n}, Y^{n}|W}(x^{n},y^{n}|w)e^{t(\sum_{k=1}^{n} \lambda(x_k,y_k)) +\frac{2t^2}{1+2tP} (nP - \sum_{k=1}^{n}x_k^2)}
  \, \mathrm{d}y^n \, \mathrm{d}x^n \notag\\
  & = {(1+2tP)}^{-\frac{1}{2}}\int_{x^{n-1}, y^{n-1}} p_{X^{n-1}, Y^{n-1}|W}(x^{n-1},y^{n-1}|w)e^{t(\sum_{k=1}^{n-1} \lambda(x_k,y_k)) +\frac{2t^2}{1+2tP} (n P - \sum_{k=1}^{n-1}x_k^2)}
  \, \mathrm{d}y^{n-1} \, \mathrm{d}x^{n-1} \notag\\
  & ={(1+2tP)}^{-1}\int_{x^{n-2}, y^{n-2}} p_{X^{n-2}, Y^{n-2}|W}(x^{n-2},y^{n-2}|w)e^{t(\sum_{k=1}^{n-2} \lambda(x_k,y_k)) +\frac{2t^2}{1+2tP} ( nP - \sum_{k=1}^{n-2}x_k^2)}
  \, \mathrm{d}y^{n-2} \, \mathrm{d}x^{n-2} \notag\\
  &\:\:\: \vdots\notag\\
 & = {(1+2tP)}^{-\frac{n-1}{2}} \int_{x_1, y_1} p_{X_1, Y_1|W}(x_1,y_1|w) e^{t\lambda(x_1,y_1) +\frac{2t^2}{1+2tP} (nP - x_1^2)}
  \, \mathrm{d}y_1 \, \mathrm{d}x_1,
\label{reverseInductionStep2}
\end{align}
where the $k^{\text{th}}$ equality follows from \eqref{reverseInductionStep1} for $\ell=k-1$.
Following \eqref{reverseInductionStep2}, we consider the following chain of equalities for each $w\in \mathcal{W}$:
\begin{align*}
&\int_{x_1, y_1} p_{X_1, Y_1|W}(x_1,y_1|w) e^{t\lambda(x_1,y_1) +\frac{2t^2}{1+2tP} (nP - x_1^2)}
  \, \mathrm{d}y_1 \, \mathrm{d}x_1 \notag\\
  & \stackrel{\eqref{memorylessStatement*}}{=} \int_{x_1} p_{X_1|W}(x_1|w) e^{\frac{2t^2}{1+2tP} (nP - x_1^2)} \int_{y_1} p_{Y_1|X_1}(y_1|x_1) e^{t\lambda(x_1,y_1)}
  \, \mathrm{d}y_1 \, \mathrm{d}x_1 \notag\\
 &  \stackrel{\eqref{memorylessStatement++}}{=}  \int_{x_1} p_{X_1|W}(x_1|w) e^{\frac{2t^2}{1+2tP} (nP - x_1^2)}  \frac{1}{\sqrt{(1+2tP)}} e^{\frac{2t^2x_1^2}{1+2tP}}
  \, \mathrm{d}x_1 \notag\\
  & =  \frac{1}{\sqrt{(1+2tP)}} e^{\frac{2t^2nP}{1+2tP}},
\end{align*}
which implies from \eqref{reverseInductionStep2} that
\begin{equation}
 \int_{x^{n}, y^{n}} p_{X^{n}, Y^{n}|W}(x^{n},y^{n}|w)e^{t\sum_{k=1}^{n} \lambda(x_k,y_k)}
  \, \mathrm{d}y^n \, \mathrm{d}x^n  = {(1+2tP)}^{-\frac{n}{2}} e^{\frac{2t^2nP}{1+2tP}}.\label{reverseInductionStep3}
\end{equation}
Combining \eqref{eqnBHT4thChain} and \eqref{reverseInductionStep3}, we have
\begin{equation}
\E_{p_{W,X^n, Y^n}}\left[e^{t\sum_{k=1}^n \lambda(X_k,Y_k)}\right] = {(1+2tP)}^{-\frac{n}{2}} e^{\frac{2t^2nP}{1+2tP}}. \label{momentGenSumXkZk}
\end{equation}
Let $\{Z_k\}_{k=1}^n$ be $n$ independent copies of the standard Gaussian random variable. A straightforward calculation reveals that
\begin{equation}
\E_{\prod_{k=1}^n p_{Z_k}}\left[e^{t\sum_{k=1}^n (-PZ_k^2 + 2\sqrt{P}Z_k)}\right] = {(1+2tP)}^{-\frac{n}{2}} e^{\frac{2t^2nP}{1+2tP}}. \label{momentGenSumXkZk*}
\end{equation}
Therefore,
\begin{equation}
\E_{p_{W,X^n, Y^n}}\left[e^{t\sum_{k=1}^n \lambda(X_k,Y_k)}\right] = \E_{\prod_{k=1}^n p_{Z_k}}\left[e^{t\sum_{k=1}^n (-PZ_k^2 + 2\sqrt{P}Z_k)}\right] \label{momentGenSumXkZk**}
\end{equation}
 by \eqref{momentGenSumXkZk} and \eqref{momentGenSumXkZk*}, i.e., the moment generating functions of $\sum_{k=1}^n (-PZ_k^2 + 2\sqrt{P}Z_k)$ and $\sum_{k=1}^n \lambda(X_k,Y_k)$ are equal. \vspace{0.04 in} \linebreak It then follows that the probability distributions of
$\sum_{k=1}^n (-PZ_k^2 + 2\sqrt{P}Z_k)$ and $\sum_{k=1}^n \lambda(X_k,Y_k)$ are equal, which \vspace{0.04 in} \linebreak implies from \eqref{defUk} and \eqref{defLambda} that the probability distributions of $\sum_{k=1}^n\frac{\log e}{2(1+P)}(-PZ_k^2 + 2\sqrt{P}Z_k+P)$ and $\sum_{k=1}^n U_k$ are equal, which then implies from \eqref{eqnBHT6thChain} that
\begin{align}
&\log M_{\mathrm{fb}}^*(n-1, \varepsilon, P) \notag\\
&\le \log\xi_n - \log\Bigg(\Pr_{\prod_{k=1}^n p_{Z_k}} \left\{\sum_{k=1}^n\frac{\log e}{2(1+P)}(-PZ_k^2 + 2\sqrt{P}Z_k+P) < \log\xi_n - \frac{n}{2}\log(1+P) \right\}-\varepsilon \Bigg). \label{eqnBHT7thChain}
\end{align}
\subsection{Applying the  Berry-Ess\'een Theorem}
Although the remaining steps for simplifying \eqref{eqnBHT7thChain} are standard (cf.\ \cite[Theorem 74]{Pol10} and \cite[Theorem 4.4]{TanBook}), we include them for completeness. We define the mean of the random variable in \eqref{eqnBHT7thChain} as
\begin{align*}
\mu &\triangleq \E_{p_{Z_1}}\left[\frac{\log e}{2(1+P)}(-PZ_1^2 + 2\sqrt{P}Z_1+P)\right] \\
& =0,
\end{align*}
the standard deviation as
\begin{align}
\sigma & \triangleq \sqrt{\Var_{p_{Z_1}}\left[\left(\frac{\log e}{2(1+P)}(-PZ_1^2 + 2\sqrt{P}Z_1+P)\right)^2\right]} \notag\\
&=\sqrt{\frac{P(P+2)(\log e)^2}{2(1+P)^2}} \label{defSigma}
\end{align}
and the third absolute moment as
\begin{equation}
T\triangleq \E_{p_{Z_1}}\left[\left|\frac{\log e}{2(1+P)}(-PZ_1^2 + 2\sqrt{P}Z_1+P)\right|^3\right]. \label{defT}
%& = \left(\frac{\log e}{2(1+P)}\right)^3(15(P+1)^3  + 18P(P+1)^2 + 12P^2(P+1)+ 8P^3),
\end{equation}
Since
\begin{align*}
T^{1/3}& \stackrel{\eqref{defT}}{=} \left(\E_{p_{Z_1}}\left[\left|\frac{\log e}{2(1+P)}(-PZ_1^2 + 2\sqrt{P}Z_1+P)\right|^3\right]\right)^{1/3}\notag\\
& \stackrel{\text{(a)}}{\le}  \frac{\log e}{2(1+P)}\left(P\left(\E_{p_{Z_1}}\left[Z_1^6\right]\right)^{1/3}+2\sqrt{P}\left(\E_{p_{Z_1}}\left[|Z_1|^3\right]\right)^{1/3}+P\right)\notag\\
& = \frac{\log e}{2(1+P)}\left(15^{1/3}P+2(2\sqrt{2/\pi})^{1/3} \sqrt{P}+P\right)
\end{align*}
where (a) follows from the triangle inequality for the $3$-norm, it follows that $T$ is finite.
Using \eqref{defSigma} and \eqref{defT} and applying Berry-Ess\'een theorem for i.i.d.\ random variables \cite[Section XVI.5]{feller}, we have the following bound for all $n\in\mathbb{N}$:
\begin{align*}
\sup_{a\in\mathbb{R}}\left|\Pr_{\prod_{k=1}^n p_{Z_k}}\left\{\frac{1}{\sigma\sqrt{n}}\sum_{k=1}^n  \frac{\log e}{2(1+P)}( -PZ_k^2 + 2\sqrt{P}Z_k+P) \le a\right\} - \Phi(a) \right| \le \frac{T}{\sigma^3 \sqrt{n}}.
\end{align*}
This implies by choosing $a= \Phi^{-1}\left(\varepsilon +\frac{2T}{\sigma^3 \sqrt{n}}\right)$ that
\begin{align}
\Pr_{\prod_{k=1}^n p_{Z_k}}\left\{\frac{1}{\sigma\sqrt{n}}\sum_{k=1}^n  \frac{\log e}{2(1+P)}( -PZ_k^2 + 2\sqrt{P}Z_k+P) <\Phi^{-1}\left(\varepsilon + \frac{2T}{\sigma^3 \sqrt{n}}\right)\right\} > \varepsilon + \frac{T}{\sigma^3 \sqrt{n}}\,. \label{BerryEsseenSt1}
\end{align}
Following \eqref{eqnBHT7thChain} and letting
\begin{equation*}
\xi_n \triangleq \frac{n}{2}\log(1+P) + \sigma\sqrt{n}\Phi^{-1}\left(\varepsilon + \frac{2T}{\sigma^3 \sqrt{n}}\right),
\end{equation*}
we can express \eqref{eqnBHT7thChain} as
\begin{align*}
&\log M_{\mathrm{fb}}^*(n-1, \varepsilon, P) \notag\\
&\le \frac{n}{2}\log(1+P) + \sigma\sqrt{n}\Phi^{-1}\left(\varepsilon + \frac{2T}{\sigma^3 \sqrt{n}}\right) \notag\\
 & \qquad - \log\Bigg(\Pr_{\prod_{k=1}^n p_{Z_k}} \left\{\sum_{k=1}^n\frac{\log e}{2(1+P)}(-PZ_k^2 + 2\sqrt{P}Z_k+P) <\sigma\sqrt{n}\Phi^{-1}\left(\varepsilon + \frac{2T}{\sigma^3 \sqrt{n}}\right)  \right\}-\varepsilon \Bigg),
\end{align*}
which implies from \eqref{BerryEsseenSt1} that
\begin{align}
\log M_{\mathrm{fb}}^*(n-1, \varepsilon, P) < \frac{n}{2}\log(1+P) + \sigma\sqrt{n}\Phi^{-1}\left(\varepsilon + \frac{2T}{\sigma^3 \sqrt{n}}\right) - \log \frac{T}{\sigma^3 \sqrt{n}}\,. \label{eqnBHT8thChain}
\end{align}
Since
\[
\Phi^{-1}\left(\varepsilon + \frac{2T}{\sigma^3 \sqrt{n}}\right) = \Phi^{-1}(\varepsilon) + \frac{2T}{\sigma^3 \sqrt{n}} \left(\Phi^{-1}\right)^\prime(c)
\]
for some $c\in [\varepsilon, 2T/\sigma^3]$ by Taylor's theorem, it follows from \eqref{eqnBHT8thChain} that there exists some real constant
\begin{equation}
\bar \kappa \triangleq  \frac{2T}{\sigma^2}\left(\Phi^{-1}\right)^\prime(c) -\log \frac{T}{\sigma^3} \label{defBarKappa}
\end{equation}
 that does not depend on~$n$ (cf.\ \eqref{defSigma} and \eqref{defT}) such that
\begin{equation*}
\log M_{\mathrm{fb}}^*(n-1, \varepsilon, P) < \frac{n}{2}\log(1+P) + \sigma\sqrt{n}\Phi^{-1}(\varepsilon) + \frac{1}{2}\log n + \bar \kappa\,,
\end{equation*}
which implies by letting
\begin{equation}
\kappa\triangleq \bar\kappa + \frac{1}{2}\log(1+P) + \sigma\Phi^{-1}(\varepsilon) + \frac{1}{2} \label{defKappa}
\end{equation}
that
\begin{equation}
\log M_{\mathrm{fb}}^*(n-1, \varepsilon, P) < \frac{n-1}{2}\log(1+P) + \sigma\sqrt{n-1}\Phi^{-1}(\varepsilon) + \frac{1}{2}\log (n-1) + \kappa   \, \label{eqnBHT9thChain}
\end{equation}
for $n\ge 2$. Combining \eqref{defCP}, \eqref{defVP}, \eqref{defSigma}, \eqref{defT}, \eqref{defBarKappa} \eqref{defKappa} and \eqref{eqnBHT9thChain}, we have \eqref{cutsetStatement}.
\section{Parallel Gaussian Channels with Feedback} \label{sectionParallelAWGNChannel}
\subsection{Problem Setup and Main Result}
We consider the parallel Gaussian channels with feedback \cite[Section~9.4]{CoverBook} consisting of  $L$~independent  AWGN channels. Let $\mathcal{L}\triangleq \{1, 2, \ldots, L\}$ be the index set for the $L$~channels. For each $k\in\{1, 2, \ldots, n\}$ and each $\ell\in \mathcal{L}$ the channel law is described as follows: In time slot~$k$, the source node $\mathrm{s}$ transmits $X_{\ell,k}$ on the $\ell^{\text{th}}$ channel and the corresponding channel output denoted by $Y_{\ell,k}$ is
\[
Y_{\ell,k} = X_{\ell,k} + Z_{\ell,k},
\]
where $\{Z_{\ell,k}\}_{\substack{k \in \{1, 2, \ldots, n\},\ell \in \mathcal{L}}}$ are independent zero-mean Gaussian random variables such that the variance of $Z_{\ell,k}$ is $\sigma_\ell^2>0$. We assume that a noiseless feedback link from~the destination node $\rm d$ to  the source node $\rm s$ exists so that $(W, \{Y_\ell^{k-1}\}_{\ell\in\mathcal{L}})$ is available for encoding $\{X_{\ell,k}\}_{\ell\in \mathcal{L}}$ for each $k\in\{1, 2, \ldots, n\}$. The codewords $\{X_{\ell}^n\}_{\ell\in\mathcal{L}}$ transmitted by~$\mathrm{s}$ should satisfy the peak power constraint $\sum_{\ell=1}^L \sum_{k=1}^n X_{\ell,k}^2 \le n P$,  where $P>0$ denotes the permissible power for $(X_1^n, X_2^n, \ldots, X_L^n)$.
In other words, $\Pr\{\sum_{\ell=1}^L \sum_{k=1}^n X_{\ell,k}^2 \le n P\}=1$. An $(n, M, P)$-feedback code for the parallel Gaussian channels with feedback is defined in a similar way to Definition~\ref{defCode}. To keep notation compact, let $\boldsymbol{X}$ and $\boldsymbol{Y}$ denote the random vectors $(X_1, X_2, \ldots, X_L)$ and $(Y_1, Y_2, \ldots, Y_L)$ respectively, and let $\mathbf{x}\triangleq (x_1, x_2, \ldots, x_L)$ and $\mathbf{y}\triangleq (y_1, y_2, \ldots, y_L)$ be realizations of $\boldsymbol{X}$ and $\boldsymbol{Y}$ respectively. The parallel Gaussian  channels with feedback is characterized by the conditional probability density function $q_{\boldsymbol{Y}|\boldsymbol{X}}$ satisfying
\begin{equation}
q_{\boldsymbol{Y}|\boldsymbol{X}}(\mathbf{y}|\mathbf{x})\triangleq \prod_{\ell=1}^L \phi_{0,\sigma_\ell^2}(y_\ell-x_\ell) \label{defAWGNchannelParallel}
\end{equation}
for all $\mathbf{x}\in \mathbb{R}^L$ and $\mathbf{y}\in\mathbb{R}^L$. The formal definitions of the parallel Gaussian  channels with feedback and the corresponding $(n, M, P, \varepsilon)$-feedback code are similar  to Definitions~\ref{defAWGNchannel} and \ref{defErrorProbability} respectively, and hence they are omitted. We will use the following proposition concerning noise random variables extensively. The proof of the proposition can be established in a standard way using \eqref{defAWGNchannelParallel} and hence is omitted.
\medskip
\begin{Proposition} \label{propositionIndepNoises}
Fix any $(n, M, P, \varepsilon)$-feedback code and let $p_{W, \boldsymbol{X}^n, \boldsymbol{Y}^n, \hat W}$ denote the probability distribution induced by the code. Then, the following two statements hold for each $k\in\{1, 2, \ldots, n\}$:
\begin{enumerate}
\item[(i):]$
p_{\boldsymbol{X}^k, \boldsymbol{Y}^{k-1},\{Y_{\ell,k}-X_{\ell,k}\}_{\ell\in\mathcal{L}}} =  p_{\boldsymbol{X}^k, \boldsymbol{Y}^{k-1}} \prod_{\ell=1}^L p_{Y_{\ell,k}-X_{\ell,k}}$.
\item[(ii):]  For each $\ell\in \mathcal{L}$, $\E_{p_{W, \boldsymbol{X}^n, \boldsymbol{Y}^n}}[Y_{\ell,k}-X_{\ell,k}]=0$ and $\E_{p_{W, \boldsymbol{X}^n, \boldsymbol{Y}^n}}[(Y_{\ell,k}-X_{\ell,k})^2]=\sigma_\ell^2$.
\end{enumerate}
\end{Proposition}
\medskip
The capacity of the parallel Gaussian channels with feedback is well-known and is achieved by the optimal power allocation among the~$L$ channels obtained from the water-filling algorithm \cite[Chapter 9.4]{CoverBook}, which yields $L+1$ real numbers denoted by $\Lambda$, $P_1$, $P_2$, $\ldots$, $P_L$ that satisfy
\begin{equation}
\sum_{\ell=1}^L P_\ell = P \label{condition(i)}
\end{equation}
and
\begin{equation}
P_\ell = \max\{0,\Lambda - \sigma_\ell^2\} \label{condition(ii)}
\end{equation}
for each $\ell\in \mathcal{L}$. Recalling the definitions of $\mathrm{C}(P)$ in \eqref{defCP}, we let
\begin{equation}
\mathrm{C}_L(P) \triangleq \sum_{\ell=1}^L \mathrm{C}(P_\ell/\sigma_\ell^2) \qquad\mbox{bits per channel use} \label{defCPparallel}
\end{equation}
%and
%\begin{equation}
%\mathrm{V}_L(P) \triangleq \sum_{\ell=1}^L \mathrm{V}(P_\ell/\sigma_\ell^2) \qquad\mbox{bits$^2$ per channel use}, \label{defVPparallel}
%\end{equation}
be the capacity of the parallel Gaussian channels \cite[Chapter 9.4]{CoverBook}. The following theorem states an upper bound on the first- and second-order asymptotics for the parallel Gaussian channels with feedback.
\medskip
\begin{Theorem} \label{thmMainResultParallel}
Fix an $\varepsilon \in (0,1)$ and let
\[
M_{\mathrm{fb}}^*(n, \varepsilon, P,L) \triangleq \max\left\{M \left|\: \parbox[c]{3.1 in}{There exists an $(n, M, P, \varepsilon)$-feedback code for the parallel Gaussian channels consisting of~$L$ independent channels with noise variances $(\sigma_1^2,\ldots,\sigma_L^2)$}\right.\right\}.
\]
Recall that the values   $\Lambda$, $P_1, P_2, \ldots, P_L$ and $\mathrm{C}_L(P)$ are  determined by \eqref{condition(i)}, \eqref{condition(ii)} and \eqref{defCPparallel}.
There exists a constant $\kappa$ not depending on~$n$ such that for each $n\in \mathbb{N}$,
\begin{equation}
\log M_{\mathrm{fb}}^*(n, \varepsilon, P,L) \le n \mathrm{C}_L(P) + \kappa\sqrt{n}. \label{cutsetStatementParallel}
\end{equation}
\end{Theorem}

\subsection{Strong Converse}
It was shown in Tan and Tomamichel's work~\cite[Appendix~A]{TanTom13a} that for each $\varepsilon\in(0,1)$, there exists a constant $\hat \kappa$  (not depending on~$n$) such that
\begin{equation}
\log M_{\mathrm{fb}}^*(n, \varepsilon, P,L) \ge n \mathrm{C}_L(P) + \sqrt{n\mathrm{V}_L(P)}\Phi^{-1}(\varepsilon) +  \frac{1}{2}\log n + \hat \kappa, \label{cutsetStatementParallelAch}
\end{equation}
where $\mathrm{C}_L(P)$ was defined in \eqref{defCPparallel} and
\begin{equation}
\mathrm{V}_L(P) \triangleq \sum_{\ell=1}^L \mathrm{V}(P_\ell/\sigma_\ell^2) \qquad\mbox{bits$^2$ per channel use} \label{defVPparallel}
\end{equation}
denotes the dispersion of the parallel Gaussian channels without feedback (first proved by  Polyanskiy~\cite[Theorem~78]{Pol10}).
%. The dispersion of this class of channels (without the $\frac{1}{2}\log n$ third-order term) was .
Theorem~\ref{thmMainResultParallel} together with \eqref{cutsetStatementParallelAch} imply that
\begin{equation}
\lim_{n\to\infty}\frac{1}{n}\log M_{\mathrm{fb}}^*(n, \varepsilon, P,L)=\mathrm{C}_L(P)
\end{equation}
for all $\varepsilon\in (0,1)$. Since the limit of the normalized logarithm of the code sizes exists and  does not depend on $\varepsilon \in (0,1)$, the strong converse is established for the parallel Gaussian channels with feedback.

\subsection{Proof of Theorem \ref{thmMainResultParallel}}
\begin{IEEEproof}
Fix an $\varepsilon\in (0,1)$ and choose an arbitrary sequence of $(\bar n, M_{\mathrm{fb}}^*(\bar n, \varepsilon, P,L), P, \varepsilon)$-feedback codes for the parallel Gaussian channels with feedback. Letting $n=\bar n+1$ and following similar procedures for proving \eqref{eqnBHTSecondChain} at the start of the proof of Theorem~\ref{thmMainResult}, we obtain a sequence of $(n, M_{\mathrm{fb}}^*(n-1, \varepsilon, P,L), P, \varepsilon)$-feedback codes with
 \begin{equation}
 \Pr\left\{\sum_{\ell=1}^L \sum_{k=1}^n X_{\ell,k}^2 = n P\right\}=1 \label{powerEqualityParallel}
 \end{equation}
 such that the following inequality holds for each $n\in \{2, 3, \ldots\}$ and each $\xi_n >0$:
\begin{align}
&\log M_{\mathrm{fb}}^*(n-1, \varepsilon, P,L)\notag\\
& \le \log\xi_n - \log\left(\Pr_{p_{W, \boldsymbol{X}^n, \boldsymbol{Y}^n}} \left\{\sum_{k=1}^n \log\frac{p_{\boldsymbol{Y}_k|\boldsymbol{X}_k}(\boldsymbol{Y}_{k}|\boldsymbol{X}_{k})}{s_{\boldsymbol{Y}_{k}}(\boldsymbol{Y}_{k})} < \log\xi_n \right\}-\varepsilon\right),\label{eqnBHTSecondChainParallel}
\end{align}
where $p_{W, \boldsymbol{X}^n, \boldsymbol{Y}^n}$ denotes the probability distribution induced by the $(n, M_{\mathrm{fb}}^*(n-1, \varepsilon, P,L), P, \varepsilon)$-feedback code, and $s_{\boldsymbol{Y}_{k}}$ is defined for each $k\in\{1, 2, \ldots, n\}$ as
\begin{equation}
s_{\boldsymbol{Y}_{k}}(\mathbf{y}_{k})=\prod_{\ell=1}^L \phi_{0, P_\ell + \sigma_\ell^2}(y_{\ell,k}) \label{defsYkParallel}
\end{equation}
for all $\mathbf{y}_{k}$. The channel law is
\begin{equation}
p_{\boldsymbol{Y}_k|\boldsymbol{X}_k}(\mathbf{y}_k|\mathbf{x}_k)=\prod_{\ell=1}^L \phi_{0,\sigma_\ell^2}(y_{\ell,k}-x_{\ell,k}) \label{memorylessStatement++parallel}
\end{equation}
for each $k\in\{1, 2, \ldots, n\}$. Combining \eqref{defsYkParallel} and \eqref{memorylessStatement++parallel}, we have
\begin{align}
& \log\frac{p_{\boldsymbol{Y}_k|\boldsymbol{X}_k}(\boldsymbol{Y}_k|\boldsymbol{X}_k)}{s_{\boldsymbol{Y}_k}(\boldsymbol{Y}_k)} \notag\\
& = \sum_{\ell=1}^L \left( \frac{1}{2}\log\left(1+\frac{P_\ell}{\sigma_\ell^2}\right)+ \frac{\log e}{2(\sigma_\ell^2+P_\ell)}\left(\frac{-P_\ell}{\sigma_\ell^2}(Y_{\ell,k}-X_{\ell,k})^2 + X_{\ell,k}^2 + 2X_{\ell,k}(Y_{\ell,k}-X_{\ell,k})\right)\right)\notag\\
&\stackrel{\text{(a)}}{=}\sum_{\ell=1}^L \left( \frac{1}{2}\log\left(1+\frac{P_\ell}{\sigma_\ell^2}\right)+ \frac{\log e}{2\Lambda}\left(\frac{-P_\ell}{\sigma_\ell^2}(Y_{\ell,k}-X_{\ell,k})^2 + X_{\ell,k}^2 + 2X_{\ell,k}(Y_{\ell,k}-X_{\ell,k})\right)\right) \label{defLogLikelihoodParallel}
\end{align}
for each $k\in\{1, 2, \ldots, n\}$, where (a) follows from \eqref{condition(ii)}. Following \eqref{defLogLikelihoodParallel}, we define the function $\lambda: \mathbb{R}^4 \rightarrow \mathbb{R}$ such that
\begin{equation}
\lambda(P, \sigma^2, x,y)\triangleq \frac{-P}{\sigma^2}(y-x)^2 + P + 2x(y-x) \label{defLambdaEllParallel}
\end{equation}
and let
\begin{equation}
U_{k} \triangleq \sum_{\ell=1}^L \lambda(P_\ell, \sigma_\ell^2, X_{\ell,k},Y_{\ell,k}) \label{defUkParallel}
\end{equation}
for each $k\in\{1, 2, \ldots, n\}$. It then follows from \eqref{powerEqualityParallel}, \eqref{defLambdaEllParallel} and \eqref{defUkParallel} that
\begin{equation*}
\Pr_{p_{W, \boldsymbol{X}^n, \boldsymbol{Y}^n}}\left\{\sum_{k=1}^n U_{k}= \sum_{k=1}^n \sum_{\ell=1}^L  \left(\frac{-P_\ell}{\sigma_\ell^2}(Y_{\ell,k}-X_{\ell,k})^2 + X_{\ell,k}^2 + 2X_{\ell,k}(Y_{\ell,k}-X_{\ell,k})\right)\right\}=1,
\end{equation*}
which implies from \eqref{defCPparallel}, \eqref{eqnBHTSecondChainParallel} and \eqref{defLogLikelihoodParallel} that
\begin{align}
\log M_{\mathrm{fb}}^*(n-1, \varepsilon, P,L)\le \log\xi_n - \log\left(\Pr_{p_{W, \boldsymbol{X}^n, \boldsymbol{Y}^n}} \left\{\sum_{k=1}^n U_{k} < \frac{2\Lambda(\log\xi_n - n \mathrm{C}_L(P))}{\log e} \right\}-\varepsilon\right).\label{eqnBHT3rdChainParallel}
\end{align}
In the rest of the proof, we would like to use Chebyshev's inequality to bound the probability term in \eqref{eqnBHT3rdChainParallel}. To this end, we will evaluate in the following
\begin{equation}
\E_{p_{W, \boldsymbol{X}^n, \boldsymbol{Y}^n}}\left[ \sum_{k=1}^n U_k \right]\stackrel{\eqref{defUkParallel}}{=}\E_{p_{W, \boldsymbol{X}^n, \boldsymbol{Y}^n}}\left[ \sum_{k=1}^n \sum_{\ell=1}^L \lambda(P_\ell, \sigma_\ell^2, X_{\ell,k},Y_{\ell,k}) \right] \label{evaluateExpSumUk1st}
 \end{equation}
 and
 \begin{align}
 &\Var_{p_{W, \boldsymbol{X}^n, \boldsymbol{Y}^n}}\left[\sum_{k=1}^n U_k \right] \notag\\*
 & \stackrel{\eqref{defUkParallel}}{=}\Var_{p_{W, \boldsymbol{X}^n, \boldsymbol{Y}^n}}\left[ \sum_{k=1}^n \sum_{\ell=1}^L \lambda(P_\ell, \sigma_\ell^2, X_{\ell,k},Y_{\ell,k}) \right]\notag\\
 & = \sum_{k=1}^n \sum_{m=1}^n \sum_{\ell=1}^L \sum_{\ell^\prime =1}^L \Cov\left[  \lambda(P_\ell, \sigma_\ell^2, X_{\ell,k},Y_{\ell,k}), \lambda(P_{\ell^\prime}, \sigma_{\ell^\prime}^2, x_{\ell^\prime,m},y_{\ell^\prime,m}) \right] \notag\\
 & = \sum_{k=1}^n \sum_{\ell=1}^L \Var_{p_{W, \boldsymbol{X}^n, \boldsymbol{Y}^n}}\left[ \lambda(P_\ell, \sigma_\ell^2, X_{\ell,k},Y_{\ell,k}) \right] + \sum_{(m, \ell^\prime)\ne (k, \ell)}\Cov\left[  \lambda(P_\ell, \sigma_\ell^2, X_{\ell,k},Y_{\ell,k}), \lambda(P_{\ell^\prime}, \sigma_{\ell^\prime}^2, x_{\ell^\prime,m},y_{\ell^\prime,m}) \right].   \label{evaluateVarSumUk1st}
  \end{align}
Following \eqref{evaluateExpSumUk1st}, we consider the following chain of equalities for each $k\in\{1, 2, \ldots, n\}$ and each $\ell\in \mathcal{L}$:
\begin{align}
& \E_{p_{W, \boldsymbol{X}^n, \boldsymbol{Y}^n}}\left[\lambda(P_\ell,\sigma_\ell^2,X_{\ell,k},Y_{\ell,k})\right] \notag\\
& \stackrel{\eqref{defLambdaEllParallel}}{=}  \E_{p_{W, \boldsymbol{X}^n, \boldsymbol{Y}^n}}\left[\frac{-P_\ell}{\sigma_\ell^2}(Y_{\ell,k}-X_{\ell,k})^2 + P_\ell + 2X_{\ell,k}(Y_{\ell,k}-X_{\ell,k})\right] \notag\\
& \stackrel{\text{(a)}}{=} \E_{p_{W, \boldsymbol{X}^n, \boldsymbol{Y}^n}}\left[ 2X_{\ell,k}(Y_{\ell,k}-X_{\ell,k})\right] \notag\\
& \stackrel{\text{(b)}}{=} 2  \E_{p_{W, \boldsymbol{X}^n, \boldsymbol{Y}^n}}\left[ X_{\ell,k}\right]\E_{p_{W, \boldsymbol{X}^n, \boldsymbol{Y}^n}}\left[Y_{\ell,k}-X_{\ell,k}\right] \notag\\
& \stackrel{\text{(c)}}{=}0, \label{evaluateExpSumUk2nd}
\end{align}
where
\begin{enumerate}
\item[(a)] follow from Statement (ii) in Proposition~\ref{propositionIndepNoises}.
%\eqref{memorylessStatement++parallel} that $(Y_{\ell,k}-X_{\ell,k})$ is a Gaussian random variable whose variance is $\sigma_\ell^2$.
\item[(b)] follows from Statement (i) in Proposition~\ref{propositionIndepNoises}.
%the fact that $X_{\ell,k}$ and $(Y_{\ell,k}-X_{\ell,k})$ are independent.
\item[(c)] follows from Statement (ii) in Proposition~\ref{propositionIndepNoises}.
%\eqref{memorylessStatement++parallel} that $(Y_{\ell,k}-X_{\ell,k})$ is a zero-mean Gaussian random variable.
\end{enumerate}
Combining \eqref{evaluateExpSumUk1st} and \eqref{evaluateExpSumUk2nd}, we have
 \begin{equation}
\E_{p_{W, \boldsymbol{X}^n, \boldsymbol{Y}^n}}\left[ \sum_{k=1}^n U_k \right] = 0. \label{expectationUkParallel}
\end{equation}
In addition, following \eqref{evaluateVarSumUk1st},
 we consider the following chain of equalities for each $\ell$, $\ell^\prime$, $k$ and $m$ such that $(\ell^\prime, m) \ne (\ell, k)$:
\begin{align}
& \E_{p_{W, \boldsymbol{X}^n, \boldsymbol{Y}^n}}\Bigg[\lambda(P_\ell,\sigma_\ell^2,X_{\ell,k} ,Y_{\ell,k}) \lambda(P_{\ell^\prime},\sigma_{\ell^\prime}^2,X_{\ell^\prime,m} ,Y_{\ell^\prime,m}) \Bigg] \notag\\
& \stackrel{\eqref{defLambdaEllParallel}}{=} \E_{p_{W, \boldsymbol{X}^n, \boldsymbol{Y}^n}}\Bigg[\Bigg(\frac{-P_\ell}{\sigma_\ell^2}(Y_{\ell,k}-X_{\ell,k})^2 + P_\ell + 2X_{\ell,k}(Y_{\ell,k}-X_{\ell,k})\Bigg) \notag\\
 &\qquad \qquad \times \Bigg(\frac{-P_{\ell^\prime}}{\sigma_{\ell^\prime}^2}(Y_{\ell^\prime,m}-X_{\ell^\prime,m})^2 + P_{\ell^\prime} + 2X_{\ell^\prime,m}(Y_{\ell^\prime,m}-X_{\ell^\prime,m})\Bigg)\Bigg] \notag\\
 & \stackrel{\text{(a)}}{=}   \E_{p_{W, \boldsymbol{X}^n, \boldsymbol{Y}^n}}\Bigg[2X_{\ell,k}(Y_{\ell,k}-X_{\ell,k})\left(\frac{-P_{\ell^\prime}}{\sigma_{\ell^\prime}^2}(Y_{\ell^\prime,m}-X_{\ell^\prime,m})^2 + P_{\ell^\prime}\right) \notag\\
 &\qquad \qquad + 2X_{\ell^\prime,m}(Y_{\ell^\prime,m}-X_{\ell^\prime,m})\left(\frac{-P_\ell}{\sigma_\ell^2}(Y_{\ell,k}-X_{\ell,k})^2 + P_\ell\right)\notag\\
 & \qquad \qquad + 4X_{\ell,k} X_{\ell^\prime,m}(Y_{\ell,k}-X_{\ell,k})(Y_{\ell^\prime,m}-X_{\ell^\prime,m})\Bigg] \notag\\
  & \stackrel{\text{(b)}}{=} \begin{cases}  \E_{p_{W, \boldsymbol{X}^n, \boldsymbol{Y}^n}}\left[2X_{\ell,k}(Y_{\ell,k}-X_{\ell,k})\right]\E_{p_{W, \boldsymbol{X}^n, \boldsymbol{Y}^n}}\Bigg[\frac{-P_{\ell^\prime}}{\sigma_{\ell^\prime}^2}(Y_{\ell^\prime,m}-X_{\ell^\prime,m})^2 + P_{\ell^\prime} \Bigg] & \text{if $k\le m$,} \vspace{0.04 in} \\  \E_{p_{W, \boldsymbol{X}^n, \boldsymbol{Y}^n}}\left[2X_{\ell^\prime,m}(Y_{\ell^\prime,m}-X_{\ell^\prime,m})\right]\E_{p_{W, \boldsymbol{X}^n, \boldsymbol{Y}^n}}\Bigg[\frac{-P_{\ell}}{\sigma_{\ell}^2}(Y_{\ell,k}-X_{\ell,k})^2 + P_{\ell} \Bigg] & \text{if $k>m$} \end{cases} \notag\\
   & \stackrel{\text{(c)}}{=} 0, \label{ExpectationLambdakLambdamParallel}
\end{align}
where
\begin{enumerate}
\item[(a)] follows from Proposition~\ref{propositionIndepNoises} that
\[
\E_{p_{W, \boldsymbol{X}^n, \boldsymbol{Y}^n}}\left[\Bigg(\frac{-P_\ell}{\sigma_\ell^2}(Y_{\ell,k}-X_{\ell,k})^2+P_\ell\Bigg)\Bigg(\frac{-P_{\ell^\prime}}{\sigma_{\ell^\prime}^2}(Y_{\ell^\prime,m}-X_{\ell^\prime,m})^2+P_{\ell^\prime}\Bigg)\right]=0.
\]
%the fact if $(\ell^\prime, m) \ne (\ell, k)$, then $(Y_{\ell,k}-X_{\ell,k})$ and $(Y_{{\ell^\prime},m}-X_{{\ell^\prime},m})$ are two independent Gaussian random variables whose variances are $\sigma_\ell^2$ and $\sigma_{\ell^\prime}^2$ respectively.
    \item[(b)] follows from Proposition~\ref{propositionIndepNoises} that:\\
    (i) If $k\le m$, $(Y_{{\ell^\prime},m}-X_{{\ell^\prime},m})$ is a zero-mean random variable that is independent of $(X_{\ell^\prime,m},X_{\ell,k}, Y_{\ell,k})$.\\
    (ii) If $k > m$, $(Y_{\ell,k}-X_{\ell,k})$ is a zero-mean random variable that is independent of $(X_{\ell,k} ,X_{\ell^\prime,m}, Y_{\ell^\prime,m})$.
        \item[(c)] follows from Statement (ii) in Proposition~\ref{propositionIndepNoises}.
        %the fact that $(Y_{\ell,k}-X_{\ell,k})$ and $(Y_{{\ell^\prime},m}-X_{{\ell^\prime},m})$ are two zero-mean random variables whose variances are  $\sigma_{\ell}^2$ and $\sigma_{\ell^\prime}^2$ respectively.
\end{enumerate}
Combining \eqref{evaluateExpSumUk2nd} and \eqref{ExpectationLambdakLambdamParallel}, we obtain
\begin{align*}
\Cov_{p_{W, \boldsymbol{X}^n, \boldsymbol{Y}^n}}\Big[\lambda(P_\ell,\sigma_\ell^2,X_{\ell,k} ,Y_{\ell,k}), \lambda(P_{\ell^\prime},\sigma_{\ell^\prime}^2,X_{\ell^\prime,m} ,Y_{\ell^\prime,m}) \Big] =0
\end{align*}
for all $(\ell^\prime, m) \ne (\ell, k)$, which implies from \eqref{evaluateVarSumUk1st} and \eqref{evaluateExpSumUk2nd} that
\begin{align}
 \Var_{p_{W, \boldsymbol{X}^n, \boldsymbol{Y}^n}}\left[\sum_{k=1}^n U_k\right]=  \sum_{k=1}^n \sum_{\ell=1}^L \E_{p_{W, \boldsymbol{X}^n, \boldsymbol{Y}^n}}\left[ \left(\lambda(P_\ell, \sigma_\ell^2, X_{\ell,k},Y_{\ell,k})\right)^2 \right]. \label{varianceSumUkParallel}
\end{align}
Following \eqref{varianceSumUkParallel}, we consider the following chain of equalities for each $k\in\{1, 2, \ldots, n\}$ and each $\ell\in \mathcal{L}$:
\begin{align}
& \E_{p_{W, \boldsymbol{X}^n, \boldsymbol{Y}^n}}\left[ \left(\lambda(P_\ell, \sigma_\ell^2, X_{\ell,k},Y_{\ell,k})\right)^2 \right] \notag\\
& \stackrel{\eqref{defLambdaEllParallel}}{=}  \E_{p_{W, \boldsymbol{X}^n, \boldsymbol{Y}^n}}\Bigg[\Bigg(\frac{-P_\ell}{\sigma_\ell^2}(Y_{\ell,k}-X_{\ell,k})^2 + P_\ell + 2X_{\ell,k}(Y_{\ell,k}-X_{\ell,k})\Bigg)^2\Bigg] \notag\\
 & \stackrel{\text{(a)}}{=} \E_{p_{W, \boldsymbol{X}^n, \boldsymbol{Y}^n}}\Bigg[2P_\ell^2 + 4 \sigma_\ell^2X_{\ell,k}^2\Bigg] \label{ExpectationUkUkParallel}
\end{align}
where (a) follows from Proposition~\ref{propositionIndepNoises} that $(Y_{\ell,k}-X_{\ell,k})/\sigma_\ell$ is a standard Gaussian random variable independent of $X_{\ell,k}$.
It then follows from \eqref{ExpectationUkUkParallel} and \eqref{powerEqualityParallel} that
\begin{align}
4nP \min_{\ell\in\mathcal{L}}\{\sigma_\ell^2\} <\sum_{k=1}^n \sum_{\ell=1}^L \E_{p_{W, \boldsymbol{X}^n, \boldsymbol{Y}^n}}\left[ \left(\lambda(P_\ell, \sigma_\ell^2, X_{\ell,k},Y_{\ell,k})\right)^2\right] \le   2n \sum_{\ell=1}^L P_\ell^2 +  4nP\max_{\ell\in\mathcal{L}}\{\sigma_\ell^2\}. \label{eqnBHT4thChainParallel}
\end{align}
Letting
\begin{equation}
\tilde \kappa \triangleq 4\min_{\ell\in\mathcal{L}}\{\sigma_\ell^2\}P \label{defTildeKappaParallel}
\end{equation}
and
\begin{equation}
\bar \kappa \triangleq 2\sum_{\ell=1}^L P_\ell^2 + 4\max_{\ell\in\mathcal{L}}\{\sigma_\ell^2\}P \label{defBarKappaParallel}
\end{equation}
be two positive real numbers, it follows from \eqref{varianceSumUkParallel}, \eqref{eqnBHT4thChainParallel}, \eqref{defTildeKappaParallel} and \eqref{defBarKappaParallel} that
\begin{equation}
 n \tilde \kappa<\Var_{p_{W, \boldsymbol{X}^n, \boldsymbol{Y}^n}}\left[\sum_{k=1}^n U_k\right] \le n \bar \kappa. \label{eqnBHT5thChainParallel}
\end{equation}
Omitting the distribution subscripts for probability, expectation and variance and letting
\begin{equation}
\log\xi_n \triangleq n\mathrm{C}_L(P) + \frac{\log e}{2\Lambda}\sqrt{\left(\frac{2}{1-\varepsilon}\right)\Var\left[\sum_{k=1}^n U_k\right]}\, ,
\end{equation}
it follows from \eqref{eqnBHT3rdChainParallel} that
\begin{align}
&\log M_{\mathrm{fb}}^*(n-1, \varepsilon, P,L)\notag\\
&\le n\mathrm{C}_L(P) + \frac{\log e}{2\Lambda}\sqrt{\left(\frac{2}{1-\varepsilon}\right)\Var\left[\sum_{k=1}^n U_k\right]} - \log\left(1-\varepsilon - \Pr\left\{\sum_{k=1}^n U_{k} \ge \sqrt{\left(\frac{2}{1-\varepsilon}\right)\Var\left[\sum_{k=1}^n U_k\right]}\right\}\right). \label{eqnBHT6thChainParallel}
\end{align}
Since $\sqrt{\left(\frac{2}{1-\varepsilon}\right)\Var\left[\sum_{k=1}^n U_k\right]}>0$ by \eqref{eqnBHT5thChainParallel}, it follows from Chebyshev's inequality that
\begin{align*}
 \Pr\left\{\sum_{k=1}^n U_{k} \ge \sqrt{\left(\frac{2}{1-\varepsilon}\right)\Var\left[\sum_{k=1}^n U_k\right]}\right\} \le (1-\varepsilon)/2,
\end{align*}
which implies from \eqref{eqnBHT6thChainParallel} that
\begin{align}
\log M_{\mathrm{fb}}^*(n-1, \varepsilon, P ,L)\le n\mathrm{C}_L(P) + \frac{\log e}{2\Lambda}\sqrt{\left(\frac{2}{1-\varepsilon}\right)\Var\left[\sum_{k=1}^n U_k\right]} - \log\left(\frac{1-\varepsilon}{2}\right). \label{eqnBHT7thChainParallel}
\end{align}
Define
\begin{equation}
\kappa\triangleq \frac{\log e}{\Lambda}\sqrt{\frac{2\bar \kappa}{1-\varepsilon}} - \log\left(\frac{1-\varepsilon}{2}\right)+ \mathrm{C}_L(P) \label{defKappaParallel}
\end{equation}
and continue the inequality in \eqref{eqnBHT7thChainParallel} for $n\ge 2$ as follows:
\begin{align}
&n\mathrm{C}_L(P) + \frac{\log e}{2\Lambda}\sqrt{\left(\frac{2}{1-\varepsilon}\right)\Var\left[\sum_{k=1}^n U_k\right]} - \log\left(\frac{1-\varepsilon}{2}\right) \notag\\
&\stackrel{\eqref{eqnBHT5thChainParallel}}{\le}  n\mathrm{C}_L(P) + \frac{\log e}{2\Lambda}\sqrt{\frac{2\bar \kappa n}{1-\varepsilon}  } - \log\left(\frac{1-\varepsilon}{2}\right) \notag\\
& \stackrel{\text{(a)}}{\le} n\mathrm{C}_L(P) + (\sqrt{n-1}+1)\frac{\log e}{2\Lambda}\sqrt{\frac{2\bar \kappa}{1-\varepsilon} } - \log\left(\frac{1-\varepsilon}{2}\right) \notag\\
&\le (n-1)\mathrm{C}_L(P) + \sqrt{n-1}\left(\frac{\log e}{\Lambda}\sqrt{\frac{2\bar \kappa}{1-\varepsilon}} - \log\left(\frac{1-\varepsilon}{2}\right) + \mathrm{C}_L(P)\right) \notag\\
& \stackrel{\eqref{defKappaParallel}}{=} (n-1)\mathrm{C}_L(P) + \kappa\sqrt{n-1}, \label{eqnBHT8thChainParallel}
\end{align}
where
\begin{enumerate}
\item[(a)] follows from the fact that $\sqrt{n} \le \sqrt{n-1}+1$.
\item[(b)] follows from our assumption $n\ge 2$ that $1\le \sqrt{n-1}$.
\end{enumerate}
The theorem then follows from combining \eqref{eqnBHT7thChainParallel} and \eqref{eqnBHT8thChainParallel}.
\end{IEEEproof}
\medskip

\subsection{Difficulties in Establishing the Exact Second-Order Asymptotics}
Unlike the case for $L=1$ where we are able to provide a converse proof for \eqref{eqn:asymp_expans_fb}, we fail to obtain a matching converse  statement to  \eqref{cutsetStatementParallelAch} for $L>1$. Instead, we can only conclude from Theorem~\ref{thmMainResultParallel} and \eqref{cutsetStatementParallelAch} that the second-order asymptotics in the asymptotic expansion of $\log M_{\mathrm{fb}}^*(n, \varepsilon, P,L)$ increases at a rate no faster than $\sqrt{n}$ (which is good enough for the purpose of the strong converse). The difficulty in obtaining a matching converse statement to \eqref{cutsetStatementParallelAch} for $L>1$ can be roughly explained as follows: For $L=1$, we can always assume without loss of generality that $\sigma_1^2=1$ and $P=P_1$, and the key equation to proving the reverse statement of \eqref{cutsetStatementParallelAch} is \eqref{powerConstraintCharacteristicFunction}, which enables the insertion of
\begin{equation}
e^{\frac{2t^2}{1+2tP} \left(nP - \sum\limits_{k=1}^{n}x_k^2\right)} \label{sumXk2=0whenL=1}
\end{equation}
in the third equality of \eqref{eqnBHT4thChain} and the cancellation of $e^{\frac{2t^2x_{n-\ell}^2}{1+2tP}}$ in the last step of \eqref{reverseInductionStep1}. Unfortunately for $L>1$, to prove the converse statement to~\eqref{cutsetStatementParallelAch}, it appears to be necessary to ensure that the following is true:
%the $e^{\frac{2t^2\sigma^2}{1+2tP} \left(nP - \sum\limits_{k=1}^{n}x_k^2\right)}$
%if we want to imitate the insertion of $e^{\frac{2t^2\sigma^2}{1+2tP} \left(nP - \sum\limits_{k=1}^{n}x_k^2\right)}$ in order to prove the converse statement of \eqref{cutsetStatementParallelAch}, then we need to ensure
\begin{equation}
e^{\sum\limits_{\ell=1}^L\frac{2t^2 \sigma_\ell^2}{1+2tP_\ell}\left(nP_\ell - \sum\limits_{k=1}^n x_{\ell,k}^2\right)}=1 . \label{sumXk2=0whenL>1}
\end{equation}
This requires the following $L$ equations to hold
 \begin{equation}
p_{W,X_\ell^n, Y_\ell^n}(w,x_\ell^n, y_\ell^n) = p_{W,X_\ell^n, Y_\ell^n}(w,x_\ell^n, y_\ell^n)\mathbf{1}\left(\left\{\sum_{k=1}^{n}x_{\ell,k}^2 = nP_\ell\right\}\right),\qquad\forall\, \ell\in L.\label{powerConstraintCharacteristicFunctionParallel}
 \end{equation}
%for all $w\in\mathcal{W}$, $x_\ell^n\in \mathbb{R}^n$ and $y_\ell^n\in\mathbb{R}^n$.
%
%for all $x^n \in\mathbb{R}^n$ on the support of $p_{X^n}$, which can no longer be derived from \eqref{powerEqualityParallel}

Unfortunately, we cannot assume (without loss of generality) that~\eqref{powerConstraintCharacteristicFunctionParallel} is true in view of \eqref{powerEqualityParallel} unless $\sigma_1^2=\sigma_2^2=\ldots = \sigma_L^2$ (which is a trivial case for the parallel Gaussian channels). Essentially we cannot guarantee that $\sum_{k=1}^n X_{\ell,k}^2 = nP_\ell$ for all $\ell\in\mathcal{L}$ with probability one; we only know that the sum $\sum_{\ell\in\mathcal{L}}\sum_{k=1}^n X_{\ell,k}^2=nP$ with probability one. Since we are able to conclude~\eqref{sumXk2=0whenL=1} from \eqref{powerEqualityParallel} for $L=1$ but unable to claim~\eqref{sumXk2=0whenL>1} from \eqref{powerEqualityParallel} for $L>1$, there is thus a discrepancy in the second- and third-order asymptotics between Theorem~\ref{thmMainResult} and Theorem~\ref{thmMainResultParallel} using the current proof technique.

However, what we are able to show using the current technique is that the third-order term for the parallel Gaussian channels {\em without feedback} is upper bounded by $\frac{1}{2}\log n + O(1)$, improving on \cite[Theorem 78]{Pol10} and matching the lower bound in \cite[Appendix~A]{TanTom13a}.
 Establishing the exact second- and third-order asymptotics for the parallel Gaussian channels {\em with feedback} is an avenue for future research.
\ifCLASSOPTIONcaptionsoff
 \newpage
\fi



%\bibliographystyle{IEEEtran}
%\begin{thebibliography}{1}
%\end{thebibliography}

\subsection*{Acknowledgments}
The authors would like to thank Y\"{u}cel Altu\u{g}, Yury Polyanskiy and  Yu Xiang for useful comments concerning the difference in the decay rates of the error probabilities  under the peak and average power constraints.% in an earlier draft of this manuscript.

The authors gratefully acknowledge financial support from  the National University of Singapore (NUS) under startup grant R-263-000-A98-750/133 and  NUS Young Investigator Award   R-263-000-B37-133.

%\bibliographystyle{IEEEtran}
%\bibliography{IEEEabrv,database}

%\begin{biography}[{\includegraphics[width=1in,height=1.25in,clip,keepaspectratio]{mshell}}]{Michael Shell}
%\end{IEEEbiography}
%% insert where needed to balance the two columns on the last page with
%%\newpage


%\vfill


\end{document}